\documentclass[acmlarge]{acmart}

\usepackage{nopageno}

\usepackage{graphicx}
\usepackage[normalem]{ulem}
\usepackage{amsmath}
\usepackage{algorithm}
\usepackage{algorithmicx}
\usepackage{algpseudocode}
\usepackage{graphicx}
\usepackage{color}
\usepackage{setspace}
\usepackage{xcolor}
\usepackage{comment}
\usepackage{fancyhdr}
\usepackage{hhline}
\usepackage{array}
\usepackage{xspace}
\usepackage{url}
\usepackage{caption}
\usepackage{subfig}
\usepackage{fancyvrb}
\usepackage{siunitx}
\usepackage{booktabs}
\usepackage{pifont}
\usepackage{flushend}
\usepackage[us,12hr]{datetime}
\usepackage{dblfloatfix}
\usepackage{hyperref}

\usepackage{mathtools}
\DeclarePairedDelimiter\ceil{\lceil}{\rceil}

\sloppypar

\usepackage{multirow}

\usepackage{listings}
\definecolor{codegreen}{rgb}{0,0.4,0}
\definecolor{codegray}{rgb}{0.3,0.3,0.3}
\definecolor{codepurple}{rgb}{0.7,0,0.82}
\definecolor{backcolour}{rgb}{0.98,0.98,0.98}
\definecolor{keywordblue}{rgb}{0.1,0.1,1}

\lstdefinestyle{mystyle}{
    backgroundcolor=\color{backcolour},   
    commentstyle=\color{codegreen},
    keywordstyle=\color{keywordblue},
    numberstyle=\footnotesize\color{codegreen},
    stringstyle=\color{codepurple},
    basicstyle=\ttfamily\small,
    breakatwhitespace=false,         
    breaklines=false,                 
    captionpos=b,                    
    keepspaces=true,                 
    numbers=left,                    
    numbersep=3pt,                  
    showspaces=false,                
    showstringspaces=false,
    showtabs=false,                  
    tabsize=4,
    float=tp,
    floatplacement=tbp,
    abovecaptionskip=0pt
}

\colorlet{numbercol}{codepurple}

\lstdefinelanguage
    {ptx}
    {
    morekeywords={bra, mov, ld, set, exit, add, local, lt, eq ,
                  R0, R1, R2, R3, R4, R5, R6, R7},
    literate = *{!}{{\textcolor{numbercol}{\ttfamily !}}}1
              {@}{{\textcolor{numbercol}{\ttfamily @}}}1
              {:}{{\textcolor{numbercol}{\ttfamily :}}}1
              {;}{{\textcolor{numbercol}{\ttfamily ;}}}1
              {.}{{\textcolor{numbercol}{\ttfamily .}}}1
              {,}{{\textcolor{numbercol}{\ttfamily ,}}}1
              {]}{{\textcolor{numbercol}{\ttfamily ]}}}1
              {[}{{\textcolor{numbercol}{\ttfamily [}}}1
    }

\setcopyright{none}
\renewcommand\footnotetextcopyrightpermission[1]{}
\makeatletter
\let\@authorsaddresses\@empty
\makeatother

\begin{document}

\newcommand\mario[1]{\noindent{\color{blue} {\bf \fbox{Mario}} {\it#1}}}
\newcommand\amir[1]{\noindent{\color{violet} {\bf \fbox{Amir}} {\it#1}}}
\newcommand\mohammad[1]{\noindent{\color{red} {\bf \fbox{MS}} {\it#1}}}
\newcommand\rachata[1]{\noindent{\color{green} {\bf \fbox{RA}} {\it#1}}}
\newcommand\TODO[1]{\noindent{\color{magenta} {\bf \fbox{TODO}} {\it#1}}}

\newcommand\ali[1]{\noindent{\color{blue} {\bf \fbox{AH}} {\it#1}}}

\newcommand{\versionnum}[0]{4.7}
\newboolean{publicversion}
\setboolean{publicversion}{true}
\definecolor{TrueBlue}{RGB}{0,0,255}

\ifthenelse{\boolean{publicversion}}{
\newcommand\newtext[0]{}
\newcommand\newtextI[0]{}
\newcommand\newtextII[0]{}
\newcommand\newtextIII[0]{}
\newcommand\newtextIV[0]{}
\newcommand\newtextV[0]{}
\newcommand\newtextVI[0]{}
\newcommand\newtextVII[0]{}
\newcommand\newtextVIII[0]{}
\newcommand\minorrev[0]{}
\newcommand\camrev[1]{\textcolor{blue} {#1}}
\newcommand\onurrev[1]{\textcolor{red} {#1}}
\newcommand\onurrevtwo[1]{\textcolor{green} {#1}}
}
{
\newcommand\newtext[0]{}
\newcommand\newtextI[0]{}
\newcommand\newtextII[0]{}
\newcommand\newtextIII[0]{}
\newcommand\newtextIV[0]{}
\newcommand\newtextV[0]{}
\newcommand\newtextVI[0]{}
\newcommand\newtextVII[0]{}
\newcommand\minorrev[0]{}
\newcommand\camrev[1]{\textcolor{blue} {#1}}

}
\newcommand{\titleShort}[0]{LTRF\xspace}
\newcommand{\titleShortnew}[0]{LTRF$_{conf}$\xspace}

\title{Enabling High-Capacity, Latency-Tolerant, and Highly-Concurrent
GPU Register Files via Software/Hardware Cooperation}

\author{Mohammad Sadrosadati}
\affiliation{\institution{Institute for Research in Fundamental Sciences (IPM)}}
\email{m.sadr89@gmail.com}
\author{Amirhossein Mirhosseini}
\affiliation{\institution{University of Michigan}}
\author{Ali Hajiabadi}
\affiliation{\institution{Sharif University of Technology}}
\author{Seyed Borna Ehsani}
\affiliation{\institution{Sharif University of Technology}}
\author{Hajar Falahati}\affiliation{\institution{Institute for Research in Fundamental Sciences (IPM)}}
\author{Hamid Sarbazi-Azad}\affiliation{\institution{Sharif University of Technology and Institute for Research in Fundamental Sciences (IPM)}}
\author{Mario Drumond}
\affiliation{\institution{EPFL}}
\author{Babak Falsafi}
\affiliation{\institution{EPFL}}
\author{Rachata Ausavarungnirun}
\affiliation{\institution{CMU}}
\author{Onur Mutlu}
\affiliation{\institution{ETH Z{\"u}rich and CMU}}

\begin{abstract}
Graphics Processing Units (GPUs) employ large register files to accommodate all active threads and accelerate context switching. Unfortunately, register files are a scalability bottleneck for future GPUs due to long access latency, high power consumption, and large silicon area provisioning. Prior work proposes hierarchical register file to reduce the register file power consumption by caching registers in a smaller register file cache. Unfortunately, this approach does not improve register access latency due to the low hit rate in the register file cache.

In this paper, we propose the Latency-Tolerant Register File (LTRF) architecture to achieve low latency in a two-level hierarchical structure while keeping power consumption low. We observe that compile-time interval analysis enables us to divide GPU program execution into intervals with an accurate estimate of a warp's aggregate register working-set within each interval. The key idea of LTRF is to prefetch the estimated register working-set from the main register file to the register file cache under software control, at the beginning of each interval, and overlap the prefetch latency with the execution of other warps. We observe that register bank conflicts while prefetching the registers could greatly reduce the effectiveness of LTRF. Therefore, we devise a compile-time register renumbering technique to reduce the likelihood of register bank conflicts. Our experimental results show that LTRF enables high-capacity yet long-latency main GPU register files, paving the way for various optimizations. As an example optimization, we implement the main register file with emerging high-density high-latency memory technologies, enabling 8$\times$ larger capacity and improving overall GPU performance by 34\%. 
\end{abstract}

\ccsdesc[500]{Hardware~Power and energy}
\ccsdesc[300]{Computer systems organization~Single instruction, multiple data}

\keywords{GPUs, Register Files, Bank Conflicts, Register Renumbering, Latency Tolerance, Parallelism, High Performance}

\maketitle

\renewcommand{\shortauthors}{M. Sadrosadati et al.}

\section{Introduction}
Graphics Processing Units (GPUs) are commonly-used accelerators, optimizing silicon organization with dense arithmetic for data-parallel workloads. Modern GPU microarchitecture relies on managing execution resources for a large number of Single-Instruction-Multiple-Data (SIMD) threads to exploit this arithmetic density and overlap the long memory access latency with computation~\cite{cuda}. Unfortunately, the maximum parallelism in GPUs is fundamentally limited by the register file capacity as the register file must accommodate all simultaneously running threads~\cite{cache1,RF_comp,cache2,sttram1,DWM_RF,murali_RF,unified-RF}. 

GPU register files face the difficult challenge of optimizing latency, bandwidth, and power consumption, while having maximal capacity. Prior work proposes increasing the register file capacity in various ways: compression~\cite{RF_comp}, virtualization~\cite{virtual_RF1,Zorua}, or silicon technologies for high-density memory cells~\cite{sttram1,sttram2,sttram3,edram1, edram2, DWM_RF, TFET}. While such proposals increase capacity without sacrificing power consumption, they typically result in higher register access latencies.

Register file caching~\cite{cache1,cache2} is a promising approach to enhancing capacity while lowering power consumption and effective access latency. Unfortunately, existing proposals for register file caching do \emph{not} achieve high enough hit rates in the register cache due to \emph{three} key problems. First, the high degree of thread-level parallelism (TLP) in GPUs causes threads to displace each  other's registers in the cache. Second, registers house temporary values that are often renamed, which reduces temporal locality in the cache. Third, because register names are \emph{not} spatially correlated, there is no spatial locality in a register cache. Due to these reasons, register file caching is ineffective at hiding latency in GPUs (\S~\ref{sec:eval}).

Our goal is to improve the effectiveness of register file caching in GPUs. To this end, we observe that registers can be effectively prefetched into the register cache using compile-time interval analysis to hide the long access latency of the main register file. An interval is a subgraph in a program's control-flow graph that has a \emph{single} entry point. Intervals have been widely used by optimizing compilers to identify loops~\cite{hecht1977flow}. We use interval analysis and software prefetching to fetch the entire set of required registers of an interval into the register cache and thus avoid the main register file access latency during the execution of the interval. 

We propose the \emph{Latency-Tolerant Register File (\titleShort{})}, a two-level hierarchical register file that employs a low-latency/low-power first-level register file cache backed up by a high-latency/high-capacity second-level main register file. \titleShort{} uses a compiler-driven software mechanism to prefetch a warp's register working-set into the register cache at the start of an interval. By fetching \emph{all} registers in the working-set together and overlapping the prefetch latency of one warp with the execution of another, \titleShort{} hides a substantial fraction of the access latency of the main register file during the execution of the interval. 

To accelerate register prefetching operations, it is crucial to avoid main register file bank conflicts. As the main register file banks are single ported, the bank conflicts increase the prefetch latency significantly. To resolve the main register file bank conflicts, we also devise a compile-time register renumbering technique on top of \titleShort{}, while preserving the correctness of the program.

By using \titleShort{}, we enable high-capacity yet long-latency main register files, paving the way for various optimizations. As an example optimization, we implement the main register file with high-density emerging memory technologies, e.g., domain wall memory~\cite{DWM,DWM_RF,DWM-2,DWM-3,DWM-5,DWM-6,tavakkol2018enabling}, enabling 8$\times$ larger capacity and improving overall GPU performance by 31\% while reducing register file power consumption by 46\%. In contrast, the state-of-the-art register file caching schemes reduce GPU performance by 14\%, on average, if the register file is enlarged by 8$\times$, as prior designs do \emph{not} focus on tolerating the latency of the main register file.

This paper makes the following contributions:
\begin{itemize}
    \item We show that prior proposals for register file caching do \emph{not} achieve high enough hit rates to effectively hide the long latencies of large main register files (\S~\ref{sec:eval}).
 
    \item We introduce \titleShort{}, a latency-tolerant hierarchical register file design, which enables high-capacity yet long-latency main register files. The key idea is to 1) estimate the register working set of a program's execution during an interval, using compile-time interval analysis, 2) prefetch the estimated register working-set from the main register file to the register file cache under software control, at the beginning of each interval, and overlap the prefetch latency with the execution of other warps.
   
    \item We devise a compile-time register-renumbering technique to mitigate the overhead of register bank conflicts in register prefetching operations.
   
    \item Our evaluations show that an optimized version of \titleShort{}, when implemented with an 8$\times$ larger yet 6.3$\times$ slower main register file, improves overall GPU throughput by 31\%, on average (up to 86\%). \titleShort{} performance is within 5\% of an ideal 8$\times$-capacity main register file that has no latency overhead. 
    
\end{itemize}

\section{Background and Motivation}
\label{sec:back}
\begin{figure}[h!]
\centering
\includegraphics[trim=5mm 12mm 5mm 5mm,width=0.5\linewidth]{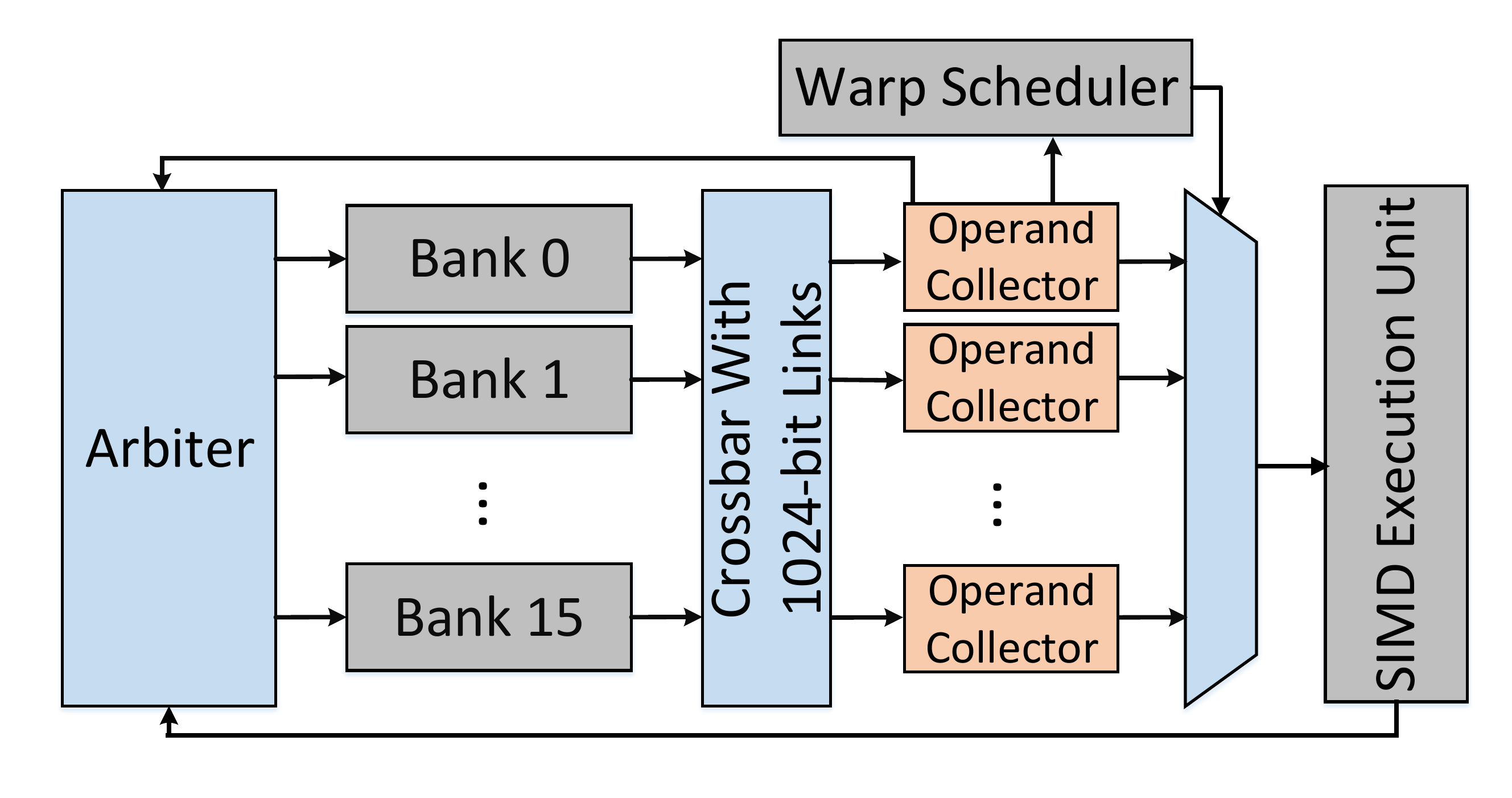}
\caption{Conventional GPU register file architecture.}
\label{fig:RF}
\end{figure}

Figure \ref{fig:RF} illustrates a conventional GPU register file architecture~\cite{RF_patent} in a streaming multiprocessor (SM). To accommodate a large number of active threads, a GPU employs a register file of megabytes in size. For example, GP100 (NVIDIA Pascal) has a register file of 14.3 megabytes in total~\cite{Nvidia_white_paper}. The register file is heavily banked (16 banks) and it allows concurrent accesses from many threads (up to 512 threads). 
Each bank stores registers from multiple warps. When the GPU issues an instruction, an operand collector concurrently accesses and gathers data associated with each thread in the issued warp's instruction through an arbiter and a large and wide crossbar, as shown in Figure~\ref{fig:RF}. The warp scheduler arbitrates among \emph{ready} warps (i.e., a warp whose operands are collected) and issues the warp's instruction to the SIMD units. 

In this section, we demonstrate the increasing demand for larger register file capacity, analyze shortcomings of prior register caching mechanisms for GPUs, and motivate the case for a design that provides high capacity without significantly increasing power consumption, on-chip die area, or access latency exposed to the GPU core.

\subsection{Factors that Limit GPU Performance}

When a warp encounters a long-latency memory instruction, the GPU selects another \emph{ready} warp to be scheduled for execution, in order to prevent the GPU core from stalling. While the applications with high TLP are more likely to contain more ready warps and are able to hide long-latency stalls more effectively, these applications with high TLP demand a large register file in order to realize their maximum TLP. To illustrate the impact of the register file size on an application's TLP, we recompile 35 workloads in CUDA SDK~\cite{GPGPU-Sim}, Rodinia~\cite{rodinia}, and Parboil~\cite{parboil} benchmark suites with the \textit{maxregcount} attribute (i.e., the attribute that enables the use of the maximum number of registers for each GPU function, i.e., 64 and 256 for Fermi and Maxwell respectively) enabled in the NVIDIA GPU compiler, nvcc. Doing so enables us to measure the number of registers applications would require if there were \emph{no} register file size constraints.

Table~\ref{tab:Extra_RF} reports the average and maximum register file capacity needed for our benchmarks to achieve the maximum TLP provided by the two GPU products. This experiment shows that a larger register file would directly translate into a larger number of executing threads, thereby increasing TLP, on average. The table corroborates our intuition that TLP is indeed limited by the number of registers and many applications benefit from compiler optimization when given a larger register file~\cite{comp-opt1,comp-opt2,comp-opt3}. The results also show that the recent version of the CUDA compiler used for Maxwell employs more aggressive compiler optimization techniques (e.g., loop unrolling) and as such enhances register usage and TLP compared to Fermi.

\begin{table}[h!]
\centering
\caption{The average and maximum register file capacity required to maximize TLP for 35 workloads in CUDA SDK~\cite{GPGPU-Sim}, Rodinia~\cite{rodinia}, and Parboil~\cite{parboil} benchmark suites in the NVIDIA Fermi and Maxwell architectures.}
\vspace{-6pt}
\begin{tabular}{|c || c |c|}
\hline 
GPU & Average required & Maximum required \\
(baseline register file size)&register file size & register file size \\ \hline \hline
Fermi (128KB)  & 184KB (1.4$\times$) & 324KB (2.5$\times$) \\ \hline
Maxwell (256KB) & 588KB (2.3$\times$) & 1504KB (5.9$\times$) \\ \hline
\end{tabular}
\label{tab:Extra_RF}
\end{table}

\subsection{Register File Scalability}
\label{sec:RF_scale}

\begin{figure}[t]
\centering
\includegraphics[trim=40mm 62mm 20mm 68mm,width=0.5\linewidth]{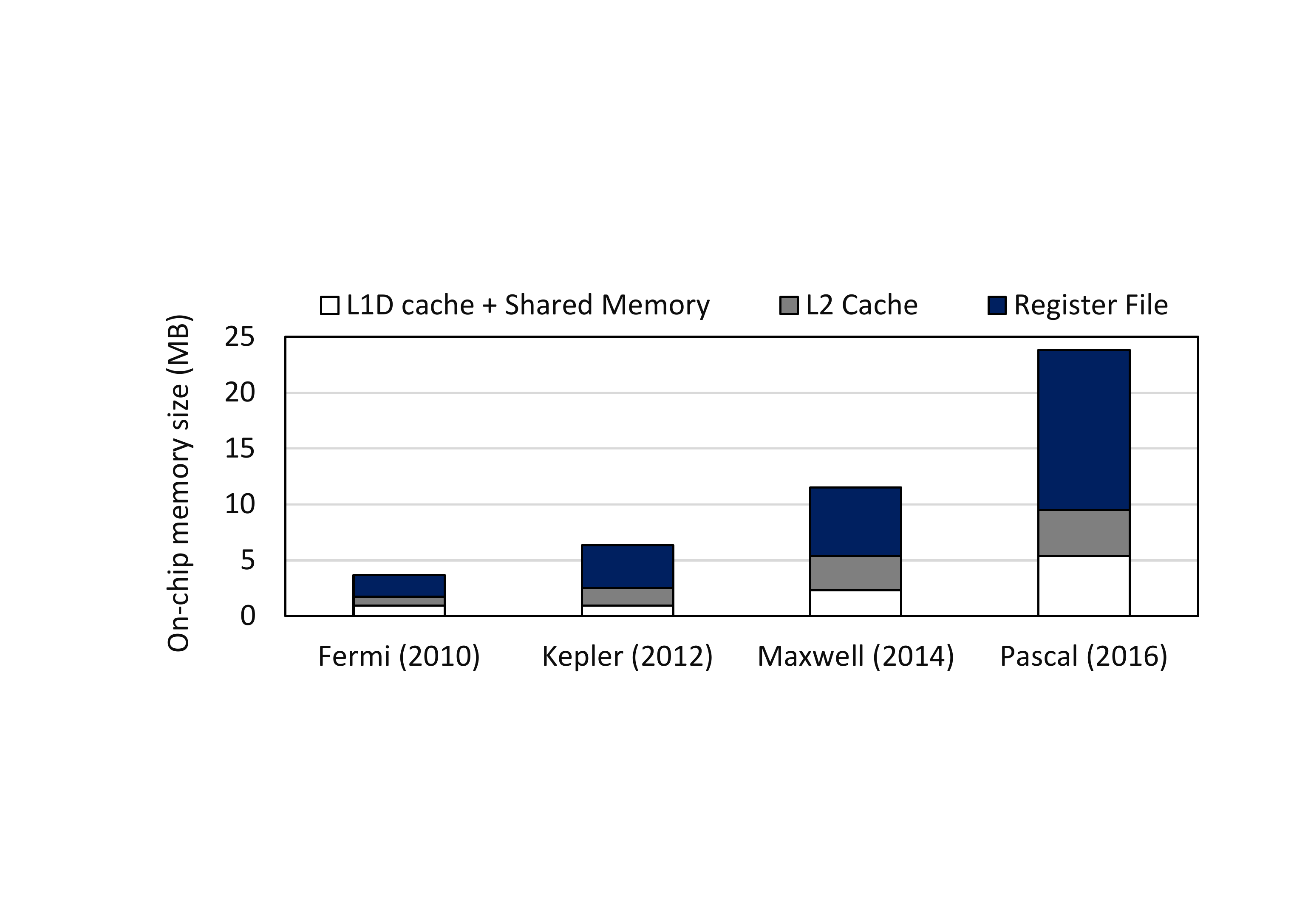}
\caption{Capacity of on-chip memory components  across generations of NVIDIA GPUs from 2010--2016.} 
\label{fig:on-chip-memory}
\end{figure}

While modern GPUs integrate more execution resources with increases in silicon density and memory bandwidth in each chip generation, the register file accounts for an increasingly larger fraction of on-chip storage, as shown in Figure~\ref{fig:on-chip-memory}. For NVIDIA Pascal~\cite{Nvidia_white_paper}, more than 60\% of the on-chip storage area, amounting to 14.3 MB is dedicated to the register file. GPU register files face the difficult challenge of optimizing latency, bandwidth, and power consumption, while having maximal capacity~\cite{sttram1, sttram2, sttram3, edram1, edram2, DWM_RF, TFET,cache1, cache2,RF_comp,samavat_2,gate1,gate3,samavat,virtual_RF1,Zorua,CABA}. Larger register files are slower, take up more silicon area and consume more power. Increasing concurrency by adding more banks exacerbates complexity and power consumption with the addition of a larger crossbar. Prior work attempts to reduce the power consumption of the register file while keeping the register access latency almost unchanged. As a result, the reduction in the power consumption is limited by the access latency of the register file. In this section, we measure the impact of various register file design parameters and configurations on register file access latency and overall GPU throughput.

\begin{table*}[h]
\footnotesize
\centering
\caption{Various register file designs with different configurations; all the numbers including number of banks ($1\times=16$), bank size ($1\times=16KB$), capacity, area, power consumption, capacity per area, capacity per power, and access latency are normalized to the baseline GPU register file with 256KB size and 16 banks.}
\vspace{-6pt}
\begin{tabular}{|c ||c |c|c| c|| c| c| c| c| c| c|}
\hline
Config. & Cell Technology &\#Banks&Bank Size&Network&Cap.&Area&Power&Cap./Area&Cap./Power&Latency\\ \hline \hline
\#1 & HP SRAM & 1$\times$ & 1$\times$ & Crossbar & 1$\times$ & 1$\times$ & 1$\times$ & 1$\times$ & 1$\times$ & 1$\times$\\ \hline
\#2 & HP SRAM & 1$\times$ & 8$\times$ & Crossbar & 8$\times$ & 8$\times$ & 8$\times$ & 1$\times$ & 1$\times$ & 1.25$\times$\\ \hline
\#3 & HP SRAM & 8$\times$ & 1$\times$ & F. Butterfly & 8$\times$ & 8$\times$ & 8$\times$ & 1$\times$ & 1$\times$ & 1.5$\times$\\ \hline
\#4 & LSTP SRAM & 1$\times$ & 8$\times$ & Crossbar & 8$\times$ & 8$\times$ & 3.2$\times$ & 1$\times$ & 2.5$\times$ & 1.6$\times$\\ \hline
\#5 & LSTP SRAM & 8$\times$ & 1$\times$ & F. Butterfly & 8$\times$ & 8$\times$ & 3.2$\times$ & 1$\times$ & 2.5$\times$ & 2.8$\times$\\ \hline
\#6 & TFET SRAM & 8$\times$ & 1$\times$ & F. Butterfly & 8$\times$ & 8$\times$ & 1.05$\times$ & 1$\times$ & 7.6$\times$ & 5.3$\times$\\ \hline
\#7 & DWM & 8$\times$ & 1$\times$ & F. Butterfly & 8$\times$ & 0.25$\times$ & 0.65$\times$ & 32$\times$ & 12$\times$ & 6.3$\times$\\ \hline
\end{tabular}

\label{tab:RF_archs}
\end{table*}

Table~\ref{tab:RF_archs} illustrates register file designs with varying parameters, including cell technology, number of banks, bank size, and network topology, relative to a baseline high performance SRAM-based design shown in Configuration \#1. The table also presents results for emerging memory cell technologies that enable a larger trade-off space between area, power and latency. We use high-performance (HP) CMOS, low-standby-power (LSTP) CMOS, tunnel-field-effect transistors (TFET), and domain-wall memory (DWM) for the cell technology~\cite{TFET_base,TFET_base_2,TFET_base_3,TFET_base_4,TFET_base_5,TFET,muralimanohar2009cacti,DWM,DWM-2,DWM-3,DWM-4,DWM-5,DWM-6,DWM-7,DWM-8,DWM-9,DWM_RF,dong2014nvsim}. To obtain these results, we first use CACTI~\cite{muralimanohar2009cacti} (the non-pipelined register file bank models) and NVSim~\cite{dong2014nvsim} to extract timing, area and power, and then feed them as parameters to GPGPU-Sim~\cite{GPGPU-Sim} to measure the average register file access latencies. The results include queuing delays incurred due to bank conflicts (our system configuration is presented in \S~\ref{sec:method}). Note that we use the flattened butterfly topology~\cite{butterfly} to reduce the overhead of the crossbar network when we increase the number of banks by 8$\times$ in our implementations. We make two key observations from Table~\ref{tab:RF_archs}. First, register file designs (such as design \#7) that minimize area and power consumption while optimizing for capacity (i.e., bits/area) exhibit higher access latency. Second, while some alternative cell technologies (e.g., DWM~\cite{DWM,DWM_RF}) can dramatically improve capacity and power consumption, they incur prohibitively long access latencies (e.g., as long as $6.3\times$ compared to the baseline register file).

\begin{figure}[t]
\begin{tabular}{c}
\includegraphics[trim=33mm 62mm 17mm 65mm,width=0.6\linewidth]{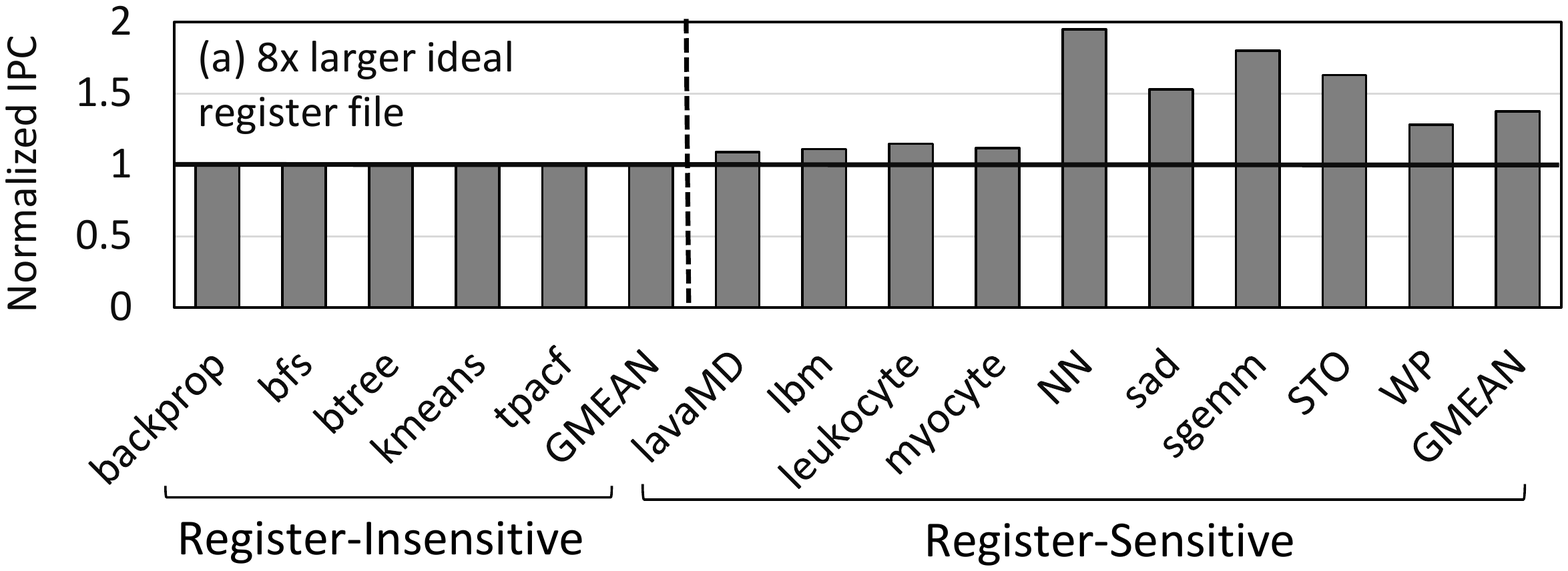}
\\
\includegraphics[trim=33mm 62mm 17mm 65mm,width=0.6\linewidth]{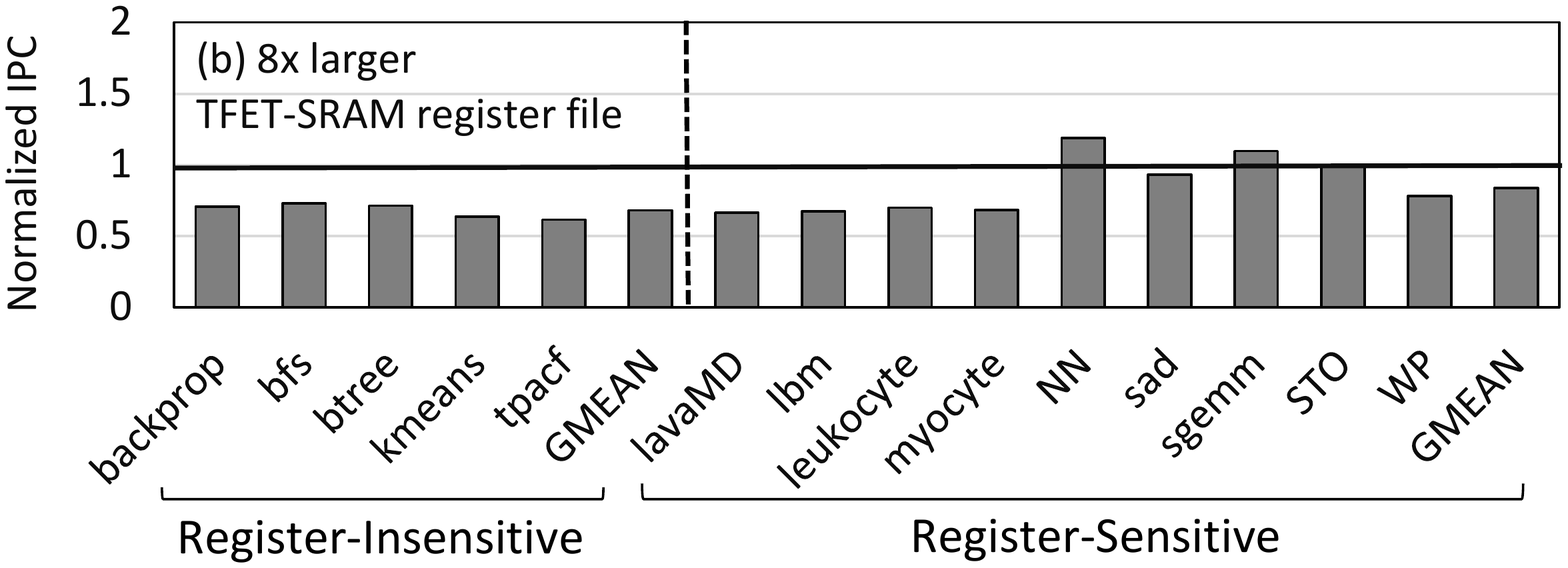}
\\
\end{tabular}
\caption{Performance effect of increasing the register file size by 8$\times$, normalized to the IPC of the baseline architecture with a 256KB register file; (a) the ideal case where the access latency of 8$\times$ larger register file is equal to the access latency of a 256KB register file, and (b) a practical solution for 8$\times$ larger register file using TFET-SRAM, that consumes power almost the same as the a 256KB baseline register files, at the price of 5.3X longer access latencies.}
\label{fig:RF_ipc}
\end{figure}

To illustrate the potential benefit of using a large register file, Figure~\ref{fig:RF_ipc}(a) plots performance (in IPC) for a high-capacity register file in the ideal case, which has the same latency as the baseline register file from Table\ref{tab:RF_archs}. We categorize our workloads into two groups: \emph{register-insensitive} and \emph{register-sensitive}. Register-insensitive workloads are the ones where the register file size is \emph{not} the bottleneck for higher TLP; i.e., increasing the register file size does \emph{not} improve TLP. We observe that increasing the register file size from 256KB to 2MB \emph{without} increasing the register file access latency, improves IPC by 10\%-95\% (37\%, on average) for register-sensitive workloads. We find that the IPC improvements are due to both more registers per thread and more warps executing in parallel. Figure~\ref{fig:RF_ipc}(b) plots performance (in IPC) for a high-capacity register file  implemented using TFET-SRAM, normalized to the baseline register file from Table~\ref{tab:RF_archs}.\footnote{We choose a 2MB TFET-SRAM register file as it consumes a similar amount of power as our baseline 256KB register file (see Table~\ref{tab:RF_archs}).} We observe that when real latencies are modeled, much of the gain from higher capacity and TLP is offset by higher latency, and overall performance \emph{reduces} despite the higher register file capacity. We conclude that register file access latency is important for performance and should be kept in check while increasing register file capacity.

\subsection{Register File Caching}
\label{sec:RFC}
\begin{figure}[h!]
\centering
\includegraphics[trim=35mm 65mm 23mm 65mm,width=0.6\linewidth]{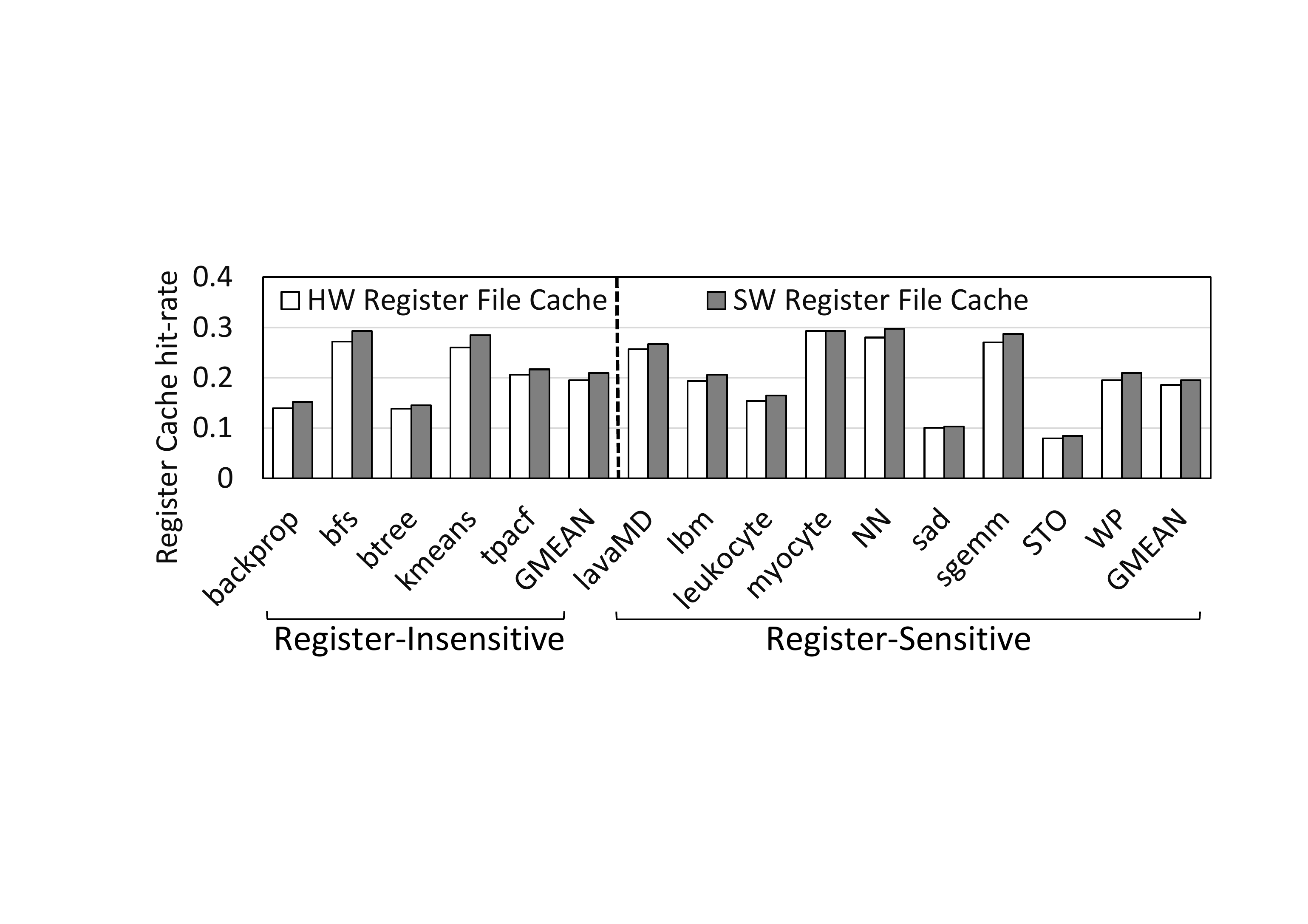}
\caption{Hit rate in hardware \cite{cache1} and software \cite{cache2} register file caches.}
\label{fig:RFC_HW_SW_hit}
\end{figure}

One method to increase the size of the register file while keeping access latency low is to cache registers in a smaller structure, i.e., register file caching. Although there is significant previous work on register file caches for CPUs~\cite{RFC_CPU2,RFC_CPU3,RFC_CPU4,RFC_CPU5,RFC_CPU6,RFC_CPU7,RFC_CPU8,RFC_CPU9,RFC_CPU10,RFC_CPU11,mirhosseini2019duplexity}, and vector processors~\cite{RFC_CPU1,RFC_CPU12,RFC_CPU13}, register file caching has \emph{not} been thoroughly investigated in GPU designs. Gebhart et al. \cite{cache1} are the first to introduce register file caches for GPUs to filter some of the accesses to the main register file and thus reduce the dynamic access energy of the main register file. The authors' design works almost the same way as a conventional cache structure and exploits temporal locality. However, as Figure~\ref{fig:RFC_HW_SW_hit} shows, for a 16KB register cache, the register cache hit rate is low: between 8\% and 30\%.

We find that the hardware register cache hit rate is low due to the following reasons: 
\begin{enumerate}
    \item Different warps can displace each other's registers in the cache due to the high warp switching rate in GPUs. This thrashing effect is also observed in SMs' local data caches~\cite{mutlu_tlp,tlp,CCWS,mirhosseini2017binochs,cache1,DYNCTA,equalizer:mahlke,sadrosadati2017effective,10.1145/3294049,micro_bakhoda,paul_gratz,pact_bakhoda,bakhoda_taco,fetch1,fetch2,fetch3,fetch4,bypass1,7753314,bypass4,bypass2,bypass3,8481514,8574555,8931588}.
    \item We find that many registers are used to only communicate results between a few instructions. As a result, these registers do \emph{not} have good temporal locality. 
    \item There is no notion of "spatial locality" in register accesses (i.e., there is no logical order among different registers).
\end{enumerate}

Follow-up work~\cite{cache2} proposes a software-managed hierarchical register file (SHRF) that aims to reduce data movement between the main register file and the register cache. However, as the main objective is to reduce dynamic energy consumption of the baseline large monolithic register file, the authors~\cite{cache2} aim to reduce the total number of accesses to the main register file, regardless of whether or not those accesses occur during the execution of a warp. 
In particular, SHRF reduces the extra register file accesses caused by register file cache write-back/reloads, by adding specialized instructions, aided by a new register allocation mechanism, to manage register movement. However, Figure~\ref{fig:RFC_HW_SW_hit} shows that the software approach does \emph{not} significantly improve the hit rate compared to a baseline hardware register cache~\cite{cache1} as it mostly focuses on reducing the number of background (i.e., write-back/reload) register accesses, rather than accesses that are needed by program instructions.

\subsection{Summary and Goals}
In this work, we leverage two observations we have provided in this section. First, the register file is one of the limiting factors in the scalability of GPUs in terms of TLP. Second, making the register file considerably larger is very difficult without sacrificing either latency or power consumption. Register caching can reduce the register access latency and thus enable aggressive power optimization techniques without degrading GPU performance. However, register caching has \emph{not} been thoroughly studied in the context of GPUs and the existing schemes mainly aim to reduce power consumption, rather than completely hide the main register file access latency. Therefore, these designs are inefficient as they do not offer high register cache hit rates. In fact, they hurt performance if used with considerably slow main register files (see \S~\ref{sec:eval}).

In this paper, we aim to architect a \emph{latency-tolerant} hierarchical register file for GPUs that can have very high capacity. Our goal is to 1) enable very high-capacity yet also high-latency main register files, while improving performance, and thus 2) open the design space for many power/area optimization techniques in the main register file that likely increase the register access latency (and thus would otherwise be unacceptable).

\section{Latency-Tolerant Register File}

To make the register file very high capacity and at the same time latency-tolerant, we propose a new register file caching mechanism that aims to 1) bring the warps' registers into the register file cache \emph{before} they are accessed by the warps (i.e., register prefetching) and 2) service all register accesses from the register file cache. As a result, the warps see the latency of a fast register cache and \emph{not} the slow main register file. We find that a near-perfect register prefetching mechanism can be implemented based on two key observations. First, the register working-set is known at compile-time as there is no indirection or aliasing in register accesses. Second, long register access latency can be hidden by the execution of other active warps.

\titleShort~takes advantage of these two observations that enable a new register prefetching scheme. \S~\ref{sec:prefetch_op} provides an overview of our register prefetching scheme. \S~\ref{sec:PR} and \S~\ref{sec:IC} provide an overview of the architectural and compiler support required for our software-driven prefetching scheme, respectively. 

\subsection{Register Prefetching Scheme}
\label{sec:prefetch_op}
We define a prefetch operation to specify which registers should be prefetched from the main register file. A prefetch operation brings the register working-set of a subgraph of the application control flow graph (CFG) into the register cache. The working-set is composed of the registers that, depending on the dynamic control flow, \emph{might be} accessed between two prefetch operations. We call subgraphs of the CFG created by prefetch operations (bounded by prefetch operations) \textit{ prefetch subgraphs}. Finding an optimal placement of prefetch operations is not only impossible in polynomial time, but also requires information available only during runtime because of dynamic warp interleavings. We propose a heuristic algorithm that employs the concept of \textit{intervals}~\cite{hecht1977flow}, subgraphs of the CFG with a single entry point, which offers compile-time analysis within a reasonable amount of time. We modify the classic interval analysis algorithm, used to find the subgraphs of the CFG with a single entry point, and introduce the concept of \emph{register-intervals} as suitable prefetch subgraphs for prefetching registers. A \emph{register-interval} is a subgraph of the CFG that 1) has a single control flow entry point and 2) requires, at most, a given number of registers. 

Our scheme brings the register working-set into the cache at the beginning of each register-interval and \emph{guarantees} that \emph{all} register accesses made inside that register-interval will be serviced from the register file cache. See \S~\ref{sec:PR} and \ref{sec:IC} for more details about the proposed register prefetching scheme.

\subsection{Architectural Support}
\label{sec:PR}
To reduce the register file cache size, we limit the number of active warps that run concurrently and maintain a pool of inactive warps; the inactive warps remain dormant and are not allocated space in the register file cache. Furthermore, we partition our register file cache and allocate each partition to an active warp, thus preventing active warps from contending for register file cache space, and thus from evicting each other's registers.  We size the dedicated caching space for each warp according to the maximum number of registers the warp can access throughout the execution of a prefetch subgraph. This parameter also sets an upper bound for the size of a prefetch subgraph working-set. By ensuring no register cache evictions occur during the execution of a prefetch subgraph, we guarantee that register movement happens only with prefetch operations or when a warp becomes active/inactive.

We deploy a two-level warp scheduler, similar to the one used in \cite{cache1,narasiman2011improving}, to schedule execution of active warps. The scheduler issues instructions from active warps in a fair manner (e.g., round-robin). Whenever a warp encounters a long latency operation, such as a data cache miss, it becomes inactive and gets replaced by another one from the active pool. The two-level scheduler enables the use of a smaller register file cache that needs to accommodate only the working-sets of the \emph{active} warps, and a warp's register working-set is swapped in and out of the register file cache as warp becomes active/inactive.

Reducing the number of \emph{active} warps provides two positive benefits: it 1) does \emph{not} limit TLP since inactive warps still maintain live state in the \emph{main} register file, and thus can be quickly activated, 2) can potentially improve performance by reducing the L1 data cache thrashing effect and by preventing \emph{all} warps from stalling at the same time~\cite{DYNCTA,equalizer:mahlke,CCWS,mutlu_tlp,narasiman2011improving}. In \titleShort{}, warp activations are not cost-free as the register working-set of the inactive warp needs to be prefetched before the warp becomes active. Hence, if we cannot hide the warp activation latency, we might negatively affect performance. In \S~\ref{subsec:ltrf-perf}, we quantitatively show that this is not the case.  \titleShort{} requires a small number of active warps to hide the warp activation latency, allowing a GPU to tolerate higher latency accesses to the main register file.\footnote{We discuss the design of the main register file and the register cache in detail in the conference version~\cite{Sadrosadati-LTRF}.} 

Prefetch operations use bit-vectors to identify the registers that should be cached for each prefetch subgraph, enabling support for various cache sizes. The prefetch bit-vector size is equal to the maximum number of registers the CUDA compiler can allocate to a thread. For example, in the latest CUDA versions, the compiler can allocate up to 256 registers to each thread, requiring a 256-bit vector for each prefetch operation. The instruction fetch unit needs to know in advance when it is going to process a prefetch bit-vector. We consider two approaches. The first embeds an extra bit in each instruction to indicate whether a prefetch bit-vector follows that instruction.  Prior work \cite{cache2} has similar requirements and the authors show that, in general, the cost of embedding the extra bit is negligible. The second approach is to add an explicit instruction that is always followed by the bit-vector.\footnote{We show in the conference version~\cite{Sadrosadati-LTRF} that code-size and performance overheads are negligible with either of the approaches.} 

When a warp becomes inactive, we must keep track of which registers should be written back and refetched once the warp becomes active again. In \titleShort{}, we simply write back and refetch the \emph{entire} register working-set of the active prefetch subgraph. 

In order to improve the efficiency of the basic \titleShort{} design, we devise operand-liveness aware \titleShort{} (called \titleShort{}+), which considers the liveness of the registers to save register file cache space. The key idea of \titleShort{}+ is to avoid writing-back/re-fetching dead registers. To this end, each read operand has to be extended with an additional bit, called the \emph{dead operand bit} as defined in~\cite{cache1}, which indicates whether the corresponding operand will be dead after the execution of the corresponding instruction. This information can be conservatively known at compile-time, using static liveness analysis. These bits are used to update the liveness bit vector. The liveness bit vector keeps track of the liveness status of all registers at the current point of execution. A register becomes live when it is written to and dead when an instruction indicates it is dead via the dead operand bit. When a warp becomes inactive, \titleShort{}+ writes back only the live registers to the main register file. When a warp becomes active, \titleShort{}+ fetches only the live registers from the main register file. \titleShort{}+ does \emph{not} read the dead registers from the main register file since their first access, if any, will be a write, and \titleShort{}+ needs to only allocate space for them in the register file cache.

\subsection{Compiler Support}
\label{sec:IC}
When a warp reaches the beginning of a \emph{prefetch subgraph}, it is paused until all of its working-set registers are loaded into the register cache. Therefore, prefetch operations may have long latencies that can potentially impose large performance overheads, and hence, they should happen infrequently. In order to address this issue, we introduce \emph{register-intervals} as effective prefetch subgraphs and partition the CFG into register-intervals. A register-interval is a subgraph of the CFG with only two constraints. First, it needs to have only one control flow entry point. Second, the number of registers used in a register-interval should \emph{not} exceed the size of a partition in the register cache.\footnote{We provide dedicated space for each active warp in the register file cache.} The primary difference between register-intervals and other similar concepts, such as \textit{strands}~\cite{cache2}, is that complex control flow structures (e.g., backward branches) are allowed inside a register-interval and they do not cause the termination of the register-interval. By relaxing such constraints, register-intervals provide two main benefits. First, register-intervals can have more static instructions and thus the number of prefetch operations can be minimized. Second, our mechanism aims to fit a loop within a single register-interval in order to increase the dynamic length of the register-intervals.

We employ classic interval analysis methods~\cite{hecht1977flow} to form register-intervals. The original interval concept~\cite{hecht1977flow}, used in classic compiler algorithms, partitions the CFG into smaller disjoint subgraphs, each with exactly one entry point. These intervals are typically used to identify loops and determine if the CFG is reducible. 
We constrain the formation algorithm to guarantee that the register working-set of each interval can fit into a register file cache partition. As a result, the register-intervals constructed by our algorithm might be smaller than the intervals formed by the original algorithm and may terminate at arbitrary points. Thus, we modified the original algorithm to construct intervals at arbitrary starting points. 

\algnewcommand\algorithmicswitch{\textbf{switch}}
\algnewcommand\algorithmiccase{\textbf{case}}
\algnewcommand\algorithmicassert{\texttt{assert}}
\algnewcommand\Assert[1]{\State \algorithmicassert(#1)}
\algnewcommand\algorithmicforeach{\textbf{for each}}
\algdef{S}[FOR]{ForEach}[1]{\algorithmicforeach\ #1\ \algorithmicdo}%
\algdef{SE}[SWITCH]{Switch}{EndSwitch}[1]{\algorithmicswitch\ #1\ \algorithmicdo}{\algorithmicend\ \algorithmicswitch}%
\algdef{SE}[CASE]{Case}{EndCase}[1]{\algorithmiccase\ #1}{\algorithmicend\ \algorithmiccase}%
\algtext*{EndSwitch}%
\algtext*{EndCase}%

\algrenewcommand\algorithmicindent{1em}

\begin{algorithm}[t]
\small
\setstretch{0.85}
\hspace*{\algorithmicindent} \textbf{Input: Application Control Flow Graph (CFG)} \\
 \hspace*{\algorithmicindent} \textbf{Output: Register-Interval CFG} 
\begin{algorithmic}[1]
\State  \textbf{Initialize:}
\ForEach {basic block : \textcolor[rgb]{0,0,0}{BB}}
\State	\textcolor[rgb]{0,0,0}{BB}.input\_list $\gets$ empty() \textcolor[rgb]{0,0.2,0.6}{// List of all register in the register cache at the begining of BB}
\State \textcolor[rgb]{0,0,0}{BB}.register-interval $\gets$ Unknown
\EndFor
\State \textcolor[rgb]{0,0,0}{Working-Set} $\gets$ empty()
\State entry\_block.register-interval $\gets$ new register-interval() \textcolor[rgb]{0,0.2,0.6}{// Each CFG has an entry basic block}
\State \textcolor[rgb]{0,0,0}{Working-Set}.insert(entry\_block)
\newline
\While{(!\textcolor[rgb]{0,0,0}{Working-Set}.empty())}
\State \textcolor[rgb]{0,0,0}{BB} $\gets$ a basic block from \textcolor[rgb]{0,0,0}{Working-Set}
\State TRAVERSE(\textcolor[rgb]{0,0,0}{BB})
\State i $\gets$ \textcolor[rgb]{0,0,0}{BB}.register-interval
\State \textbf{while} ($\exists$ basic block h for which h.register-interval==Unknown \textbf{\&} \quad\quad all of h predecessors belong to i \textbf{\&} union(output\_list of all S predecessors).size()$\leq$N) \textcolor[rgb]{0,0.2,0.6}{// N is the maximum number of registers allowed in the register-interval (i.e., size of a partition in the register file cache}) \quad \textbf{do}
\State \hspace{1em} h.register-interval $\gets$ i
\State \hspace{1em} h.input\_list $\gets$ union(output\_list of all h predecessors)
\State \hspace{1em} TRAVERSE(h)
\State \textbf{end while}
\ForEach{S $\in$ i.successors()}
\If{(S.register-interval==Unknown)}
\State S.register-interval $\gets$ new register-interval()
\State S.input\_list $\gets$ empty()
\State \textcolor[rgb]{0,0,0}{Working-Set}.insert(S)
\EndIf
\EndFor
\EndWhile
\newline
\Procedure{traverse}{\textcolor[rgb]{0,0,0}{BB}}
\State register\_list $\gets$ \textcolor[rgb]{0,0,0}{BB}.input\_list
\ForEach{instruction in \textcolor[rgb]{0,0,0}{BB}}
\State update register\_list
\If{(register\_list.size()$>$N)}
\State cut \textcolor[rgb]{0,0,0}{BB} and introduce a new basic block : \textcolor[rgb]{0,0,0}{BB1}
\State \textcolor[rgb]{0,0,0}{BB1}.register-interval $\gets$ new register-interval()
\State \textcolor[rgb]{0,0,0}{BB1}.input\_list $\gets$ empty()
\State \textcolor[rgb]{0,0,0}{Working-Set}.insert(\textcolor[rgb]{0,0,0}{BB1})
\State \textcolor[rgb]{0,0,0}{BB}.output\_list $\gets$ register\_list \textcolor[rgb]{0,0.2,0.6}{//  List of all registers in the register file cache at the end of BB}
\State exit
\EndIf
\EndFor
\EndProcedure
\end{algorithmic}
\caption{Register-Interval Formation: Pass 1.}
\label{alg:phase_one}
\end{algorithm}

Our register-interval formation algorithm is a multi-pass algorithm. Algorithm~\ref{alg:phase_one} shows the first pass. The algorithm tries to compose register-intervals with as many basic blocks as possible. Therefore, it initializes the first register-interval with the entry basic block (line 8) and iteratively attempts to add subsequent blocks to it (lines 9-25). A candidate block must satisfy two conditions to be successfully added: 1) it must be entered only from the current register-interval, 2) the register file cache space allocated for a warp must be enough to house both the active registers already in the register-interval and the ones added by the new block.  The algorithm stops when it cannot find any basic blocks that meet these conditions (line 13). After it finishes the first register-interval, it creates new register-intervals out of all the basic blocks with incoming edges from that register-interval (lines 18-24). When a register-interval is completely formed, all of the basic blocks that have incoming edges from that register-interval become new register-intervals' headers. If a single basic block's active registers do \emph{not} fit into the remaining register file cache space for that register-interval, the basic block is split across two or more register-intervals (lines 30-37). We also split the basic blocks at function calls (each function call becomes a separate register-interval). Algorithm 1 is not multi-pass and ends when all basic blocks of the control flow graph are assigned to register-intervals. After executing Algorithm 1, the CFG is transformed into a Register-Interval CFG where the nodes represent the register-intervals rather than basic blocks.

Algorithm~\ref{alg:phase_two} shows the second pass of our register-interval formation algorithm. This pass reduces the Register-Interval CFG into a smaller number of register-intervals. It works similarly to the first pass, with the difference that it never splits register-intervals. Instead, it merges two register-intervals if 1) one of them can be reached only from the other and 2) the union of their register working-sets still fits into the allocated register file cache space (lines 12-15). The second pass is repeated until the CFG can \emph{not} be reduced anymore.  After each pass of Algorithm 2, there are two possible scenarios: 1) the graph is not reduced, and 2) the graph is reduced. In the first scenario, Algorithm 2 terminates. In the second scenario, Algorithm 2 runs again on the reduced register-interval control flow graph. However, this cannot last forever, as in the worst-case (worst case in terms of algorithm execution time), the graph will be reduced to a one-node graph, which cannot be reduced more.

\begin{algorithm}[t]
\small
\setstretch{0.85}
\hspace*{\algorithmicindent} \textbf{Input: Register-Interval CFG} \\
 \hspace*{\algorithmicindent} \textbf{Output: Reduced Register-Interval CFG} 
\begin{algorithmic}[1]
\State  \textbf{Initialize:}
\ForEach {register-interval : \textcolor[rgb]{0,0,0}{i}}
\State	\textcolor[rgb]{0,0,0}{i}.register-interval $\gets$ Unknown
\EndFor
\State \textcolor[rgb]{0,0,0}{Working-Set} $\gets$ empty()
\State entry\_register-interval.next\_level\_register-interval $\gets$ new \, \, \, \, \, \, next\_level\_register-interval()
\State \textcolor[rgb]{0,0,0}{Working-Set}.insert(entry\_register-interval)
\newline
\While{(!\textcolor[rgb]{0,0,0}{Working-Set}.empty())}
\State \textcolor[rgb]{0,0,0}{i} $\gets$ a register-interval from \textcolor[rgb]{0,0,0}{Working-Set}
\State \textcolor[rgb]{0,0,0}{ii} $\gets$ \textcolor[rgb]{0,0,0}{i}.next\_level\_register-interval
\State \textcolor[rgb]{0,0,0}{ii}.register\_list $\gets$ \textcolor[rgb]{0,0,0}{i}.register\_list
\State \textbf{while} ($\exists$ register-interval h for which
         h.next\_level\_register-interval==Unknown \textbf{\&}
         all of h predecessors belong to \textcolor[rgb]{0,0,0}{ii} \textbf{\&}
         union(register\_list of all h predecessors).size()$\leq$N)
         \textcolor[rgb]{0,0.2,0.6}{// N is the maximum number of registers allowed in the register-interval (i.e., size of a partition in the
        register file cache) \quad \textbf{do}}
\State \hspace{1em} h.next\_level\_register-interval $\gets$ \textcolor[rgb]{0,0,0}{ii}
\State \hspace{1em} \textcolor[rgb]{0,0,0}{ii}.register\_list $\gets$ union(\textcolor[rgb]{0,0,0}{ii}.register\_list \& h.register\_list)
\State \textbf{end while}
\ForEach{S $\in$ \textcolor[rgb]{0,0,0}{ii}.successors()}
\If{(S.next\_level\_register-interval==Unknown)}
\State S.next\_level\_register-interval $\gets$ new next\_level\_register-interval()
\State \textcolor[rgb]{0,0,0}{Working-Set}.insert(S)
\EndIf
\EndFor
\EndWhile
\end{algorithmic}
\caption{Register-Interval Formation: Pass 2.}
\label{alg:phase_two}
\end{algorithm}

The only control flow constraint imposed by intervals is that a node can only join an interval when all of the incoming edges to the node come from that interval. As a result, backward edges and thus loop headers always create new intervals.\footnote{This is true only for reducible CFGs with natural loops where the loop has only one entry point~\cite{hecht1977flow}. However, this is usually the case as standard languages can usually only represent natural loops (except in some cases with irregular control flow structures, such as GOTO) and compiler infrastructures only produce reducible CFGs~\cite{lattner2004llvm}.} 
This key feature of intervals makes them ideal subgraphs for our purpose. By starting a new register-interval for each loop, Algorithm~\ref{alg:phase_two} maximizes the probability that an entire loop can fit into the register-interval, thereby minimizing the number of prefetch operations to one for the entire loop. 

\begin{figure}[t]
\centering
\includegraphics[trim=15mm 8mm 10mm 10mm,width=0.7\linewidth]{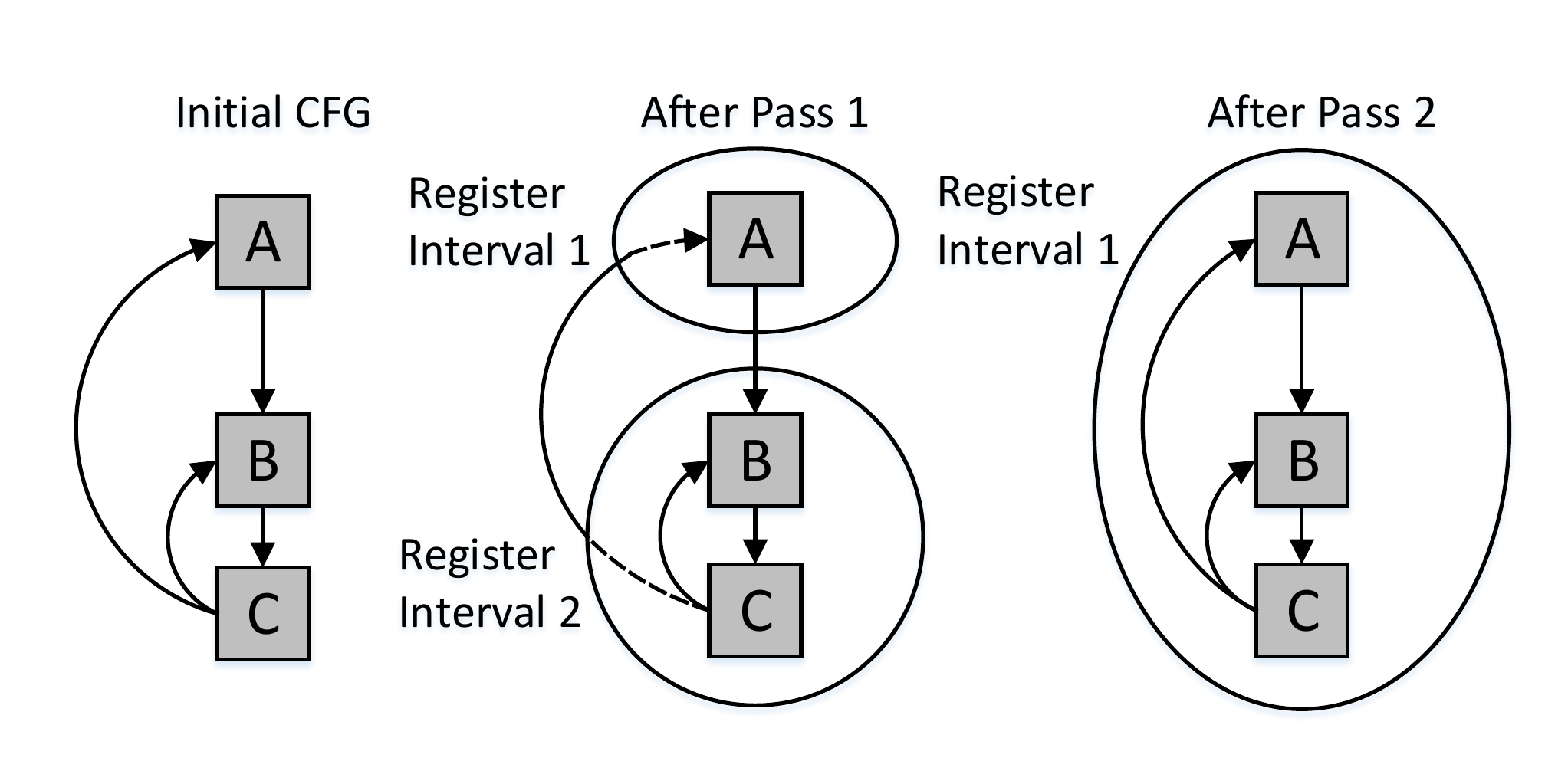}
\caption{Register-interval formation for a simple nested loop example. A,B,C represent basic blocks.}
\label{fig:interval-example}
\end{figure}

The primary role of the second pass is to prevent the mentioned control flow constraint from splitting large register-intervals into multiple smaller ones. As an example, consider the two nested loops in Figure~\ref{fig:interval-example}. 
Assuming the entire register working-set of the graph fits in the register file cache, in the first pass, basic block "A" forms register-interval 1. Basic block "B" cannot be merged with register-interval 1 as it has another incoming edge from basic block "C". Therefore, basic block "B" forms a new register-interval, named register-interval 2. Basic block "C" can be merged into register-interval 2 as basic block "C" has only one incoming edge from register-interval 2. As a result, the innermost loop becomes a separate register-interval but it cannot be merged into the outermost loop.  
In the second pass, register-interval 1 can be merged into register-interval 2 as register-interval 1 has only one incoming edge from register-interval 2. Thus, the whole outermost loop can be reduced to a single register-interval. Each repetition of the second pass of the algorithm reduces the depth of a nested loop by one if the resulting register working-set is small enough to fit in the register file cache. 

We open source the C implementation of Algorithms~\ref{alg:phase_one} and \ref{alg:phase_two} in~\cite{register-interval-github}.

\section{Resolving Register Bank Conflicts}
\label{sec:BC}

Register prefetching latency affects  \titleShort{} effectiveness because long latencies call for a higher number of active warps and larger register-cache capacities. Therefore, it is crucial to reduce the register prefetching latency in order to minimize \titleShort{}'s register-cache size. One of the main factors that greatly affects register prefetching latency is register bank conflicts during prefetching operations. When several registers of a register-interval reside in the same register bank, they cause register prefetching to happen serially, significantly increasing the register prefetching latency.

\begin{figure}[h]
\centering
\includegraphics[trim=30mm 50mm 30mm 50mm,width=0.5\linewidth]{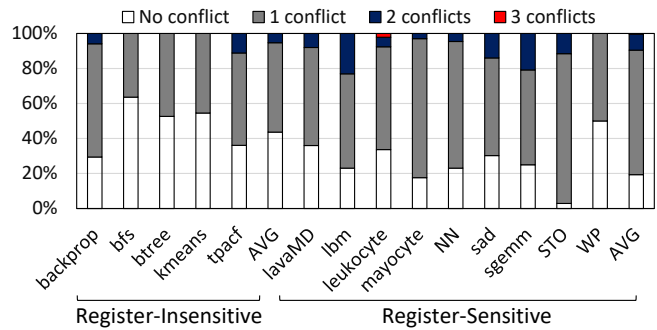}
\caption{Distribution of the number of register bank conflicts in register-intervals for different workloads.}
\label{fig:bankconflict-ltrf}
\end{figure}

To evaluate the probability of register bank conflicts while \titleShort{} prefetches registers' contents, we calculate the number of registers that reside in the same register bank in each register-interval. We assume that the maximum number of allowed registers in each register-interval is 16, and there are 16 main register file banks. A register-interval has no register bank conflict if all registers of its register working-set reside in different main register file banks. A register-interval has \emph{N} register bank conflicts if at most \emph{N+1} registers reside in the same register bank. Figure~\ref{fig:bankconflict-ltrf} shows the distribution of the number of register bank conflicts in register-intervals for different register-insensitive and register-sensitive workloads. We make two key observations. First, about 60\% and 80\% of register-intervals have at least one register bank conflict for register-insensitive and register-sensitive applications, respectively. Second, we observe up to three register bank conflicts for our workloads. We conclude that we can greatly reduce the prefetching latency by resolving register-bank conflicts.

In the rest of this section, \S~\ref{sec:ba} describes the register-bank conflict problem in more detail and provides the basic insight of our approach, which is careful register bank assignment to resolve register bank conflicts, \S~\ref{sec:rr} presents our proposed mechanism to resolve register bank conflicts in register-intervals, and \S~\ref{sec:example} elaborates on our mechanism using an example.

\subsection{Register Bank Assignment}
\label{sec:ba}
Register allocation is one of the most important optimization phases in compilers that directly affects the performance of the generated code~\cite{aho,chaitin, briggs, poletto, hack}. As the register file in a GPU is multi-banked, the register allocation mechanism should attempt to allocate the registers to different banks in order to benefit from multi-banking. Prior attempts on compilers produce bank-conflict-free register accesses by performing a register bank assignment pass in addition to register allocation \cite{zhuang, dally, lai}. Prior works build a \textit{Register Conflict Graph} whose nodes are \textit{live ranges} of the program and two nodes are connected if there is a conflict between them, e.g.,  two source operands of an instruction.

The problem we seek to solve is different. Our primary goal is to perform a bank assignment wherein all registers in the working-set of each register-interval reside in different register banks, hence decreasing the register prefetching latency. To specify our problem, it is necessary to define a key concept called \emph{register-live-range}. A register-live-range is a chain of common uses of a specific register which specifies the liveness of the register in register-intervals. The main purpose of introducing register-live-ranges is to track the liveness of values and registers across different register-intervals. In other words, register-live-ranges enable us to guarantee the correctness of the final register bank assignment with respect to the live values in registers. The input to our problem is a set of register-live-ranges specified by the register-interval CFG. Our problem is to map each register-live-range to one of register banks in such a way that those with common register-intervals reside in different register banks.

We propose a new register bank assignment technique, called \emph{register renumbering}, as a compiler optimization pass after the register allocation and register-interval formation phases. Our solution is analogous to prior bank assignment proposals. As mentioned before, register-live-ranges can be shared between different register-intervals and it is not straightforward to assign a register bank to each register-live-range. Hence, how we model our problem is important. We use graph coloring as a simplification abstraction. We model our bank assignment problem as a graph coloring problem which restricts register-live-ranges to reside in different register banks if they have a live value in the same register-intervals. The next section explains our mechanism in more detail.

\subsection{Register Renumbering Mechanism}
\label{sec:rr}
\begin{figure*}[t!]
\centering
\includegraphics[width=\linewidth]{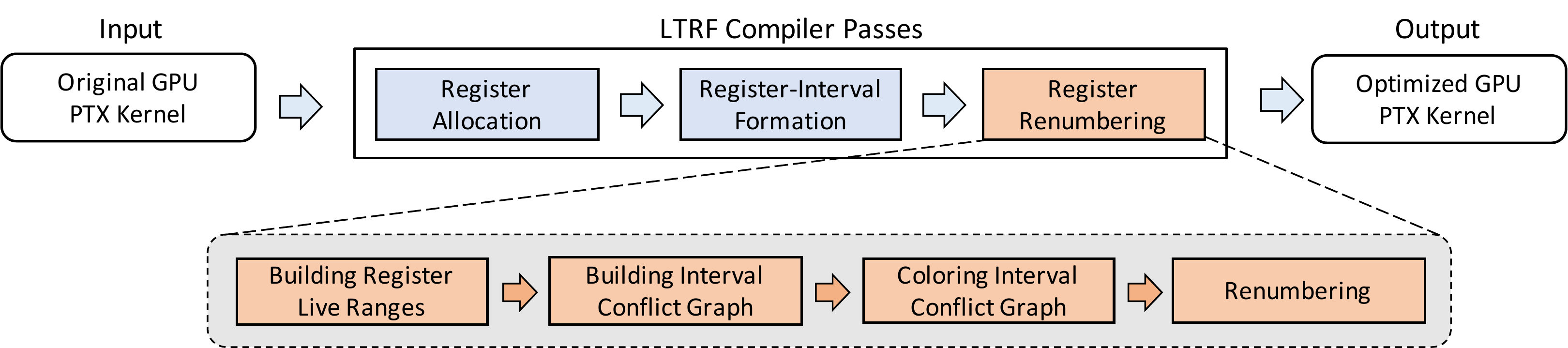}
\caption{Overview of \titleShort{} compiler support.}
\label{fig:RR}
\end{figure*}

Figure~\ref{fig:RR} presents an overview of the compiler support of \titleShort{}, including the register renumbering mechanism. As discussed before, the main goal of the register renumbering technique is to minimize register bank conflicts during prefetch operations as much as possible. This technique is employed at compile-time after register allocation and register-interval formation phases. The output of this optimization pass is register-intervals with minimal register bank conflicts. In some cases, register bank conflicts cannot be resolved unless we spill some of the register-live-ranges to main memory. We do \emph{not} produce any spill code in the register renumbering phase, since the register prefetching latency would be increased significantly due to accessing the register-live-ranges spilled in main memory.

We model our problem as a graph coloring problem. To this end, we build an appropriate graph, called the \emph{Interval Conflict Graph} (\emph{ICG}),  containing all constraints of our problem. In ICG, the nodes are register-live-ranges, and the edges represent conflicts between adjacent nodes. In other words, two nodes are connected if the two corresponding register-live-ranges have live values in same register-intervals. If we assume the number of available colors is equal to the number of register banks in a register file, a valid ICG coloring is one where all adjacent nodes have different colors, i.e., conflicting register-live-ranges receive different colors. This coloring represents our desired bank assignment with minimal register bank conflicts. 

Our register renumbering mechanism has four phases (see Figure \ref{fig:RR}):
\begin{enumerate}
    \item \textbf{Building Register-Live-Ranges}. This phase identifies register-live-ranges as the chains of definitions and uses of registers. We then identify which register-live-range is included in which register-interval(s). 
    \item \textbf{Building ICG}. Nodes of ICG are register-live-ranges and there is an edge between two nodes if they are included in at least one common register-interval.     
    \item  \textbf{Coloring ICG}. This phase colors the ICG with $N_B$ colors, where $N_B$ is the number of register banks. The problem of \textit{k-coloring} a graph is \textit{NP-complete}~\cite{cormen, kleinberg}. Therefore, we use a heuristic algorithm for graph coloring to create the colored graph in a reasonable amount of time. The heuristic algorithm that we employ is devised by Chaitin et al.~\cite{chaitin}. The algorithm requires $O(n+e)$ time, where $n$ is the number of register-liver-ranges and $e$ is the number of edges in the ICG~\cite{chaitin}. This algorithm does its best to color the graph in a balanced manner, i.e., colors are almost equally used. This key feature of this algorithm enables us to perform a \textit{balanced} bank assignment that minimizes the register bank conflicts. 
    \item \textbf{Renumbering}. The final phase of the algorithm is renumbering all registers based on their ICG color. In this phase, we assume no register has been allocated to register-live-ranges and we have a set of free registers. Whenever we want to allocate a register to a register-live-range, we choose one of the free registers of the register bank specified by the colored ICG. This approach guarantees the \emph{correctness} of the code as register-live-ranges contain all uses of a specific register. Note that each color in the colored ICG represents one of the register banks.
\end{enumerate}

\subsection{A Walk-Through Example}
\label{sec:example}
\begin{figure}[b]
	\centering
	\includegraphics[width=\linewidth]{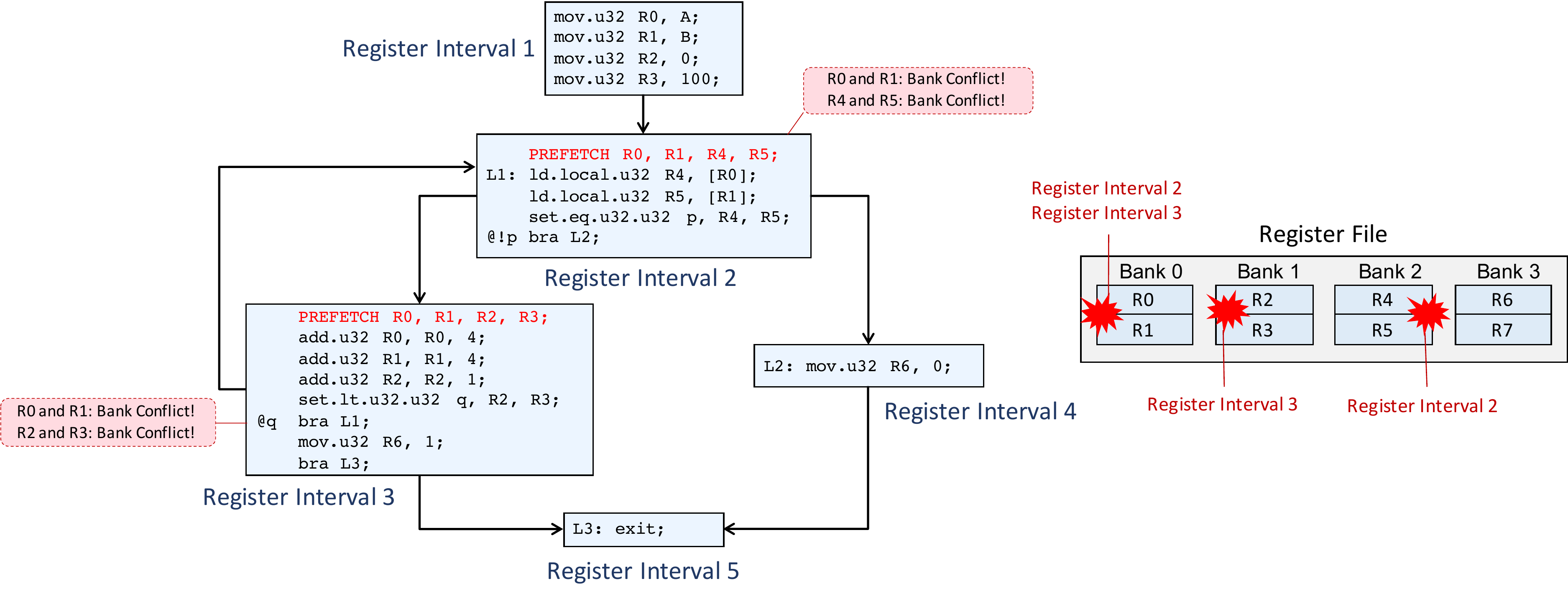}
	\caption{Register-intervals of the example PTX code in Listing~\ref{code}.}
	\label{fig:cfg-1}
\end{figure}

We elaborate on our register renumbering mechanism using a simple example. Listing~\ref{code} and Figure~\ref{fig:cfg-1} show a PTX code example and the corresponding register-interval CFG, respectively. This code example compares two arrays with 100 elements and sets a register, i.e., \texttt{R6}, if all corresponding elements of the two arrays are the same.  For the sake of simplicity, we assume that each register-interval can have up to four registers, and there are four main register banks, each containing two registers. Figure~\ref{fig:cfg-1} shows that the prefetch operation of register-interval \#2 prefetches four registers, including \texttt{R0}, \texttt{R1}, \texttt{R4}, and \texttt{R5}, which results in register bank conflicts as \texttt{R0} and \texttt{R1} are in register bank \#0 and \texttt{R4} and \texttt{R5} are in register bank \#2 (See the right side of Figure~\ref{fig:cfg-1}). As a result, performing the prefetch operation in register-interval \#2 needs two serial register bank accesses. A similar scenario happens in register-interval \#3. To resolve the register bank conflicts, we perform our register renumbering pass on this code.

\begin{figure}[t]
\begin{lstlisting}[caption=PTX Code Example, label=code]
    mov.u32         R0, A;
    mov.u32         R1, B;
    mov.u32         R2, 0;
    mov.u32         R3, 100;
L1: ld.local.u32    R4, [R0];
    ld.local.u32    R5, [R1];
    set.eq.u32.u32  p, R4, R5;
@!p bra             L2;
    add.u32         R0, R0, 4;
    add.u32         R1, R1, 4;
    add.u32         R2, R2, 1;
    set.lt.u32.u32  q, R2, R3;
@q  bra             L1;
    mov.u32         R6, 1;
    bra             L3;
L2: mov.u32         R6, 0;
L3: exit;
\end{lstlisting}
\end{figure}

\begin{figure}[h]
\centering
\includegraphics[trim=0mm 6mm 0mm 0mm,width=\linewidth]{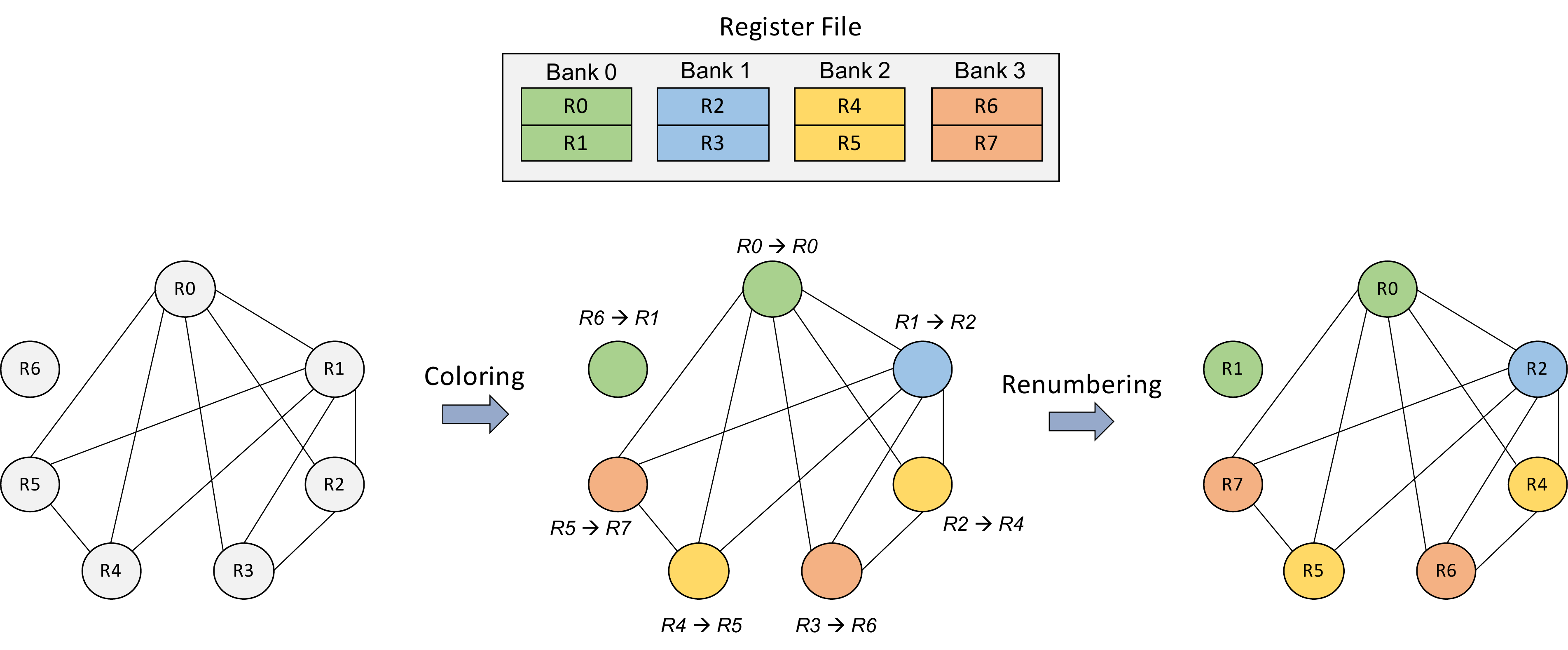}
\caption{Top: Register File, Bottom: ICG coloring and renumbering of PTX code in Listing~\ref{code}.}
\label{fig:ICG}
\end{figure}

\begin{figure}[h]
	\centering
	\includegraphics[width=\linewidth]{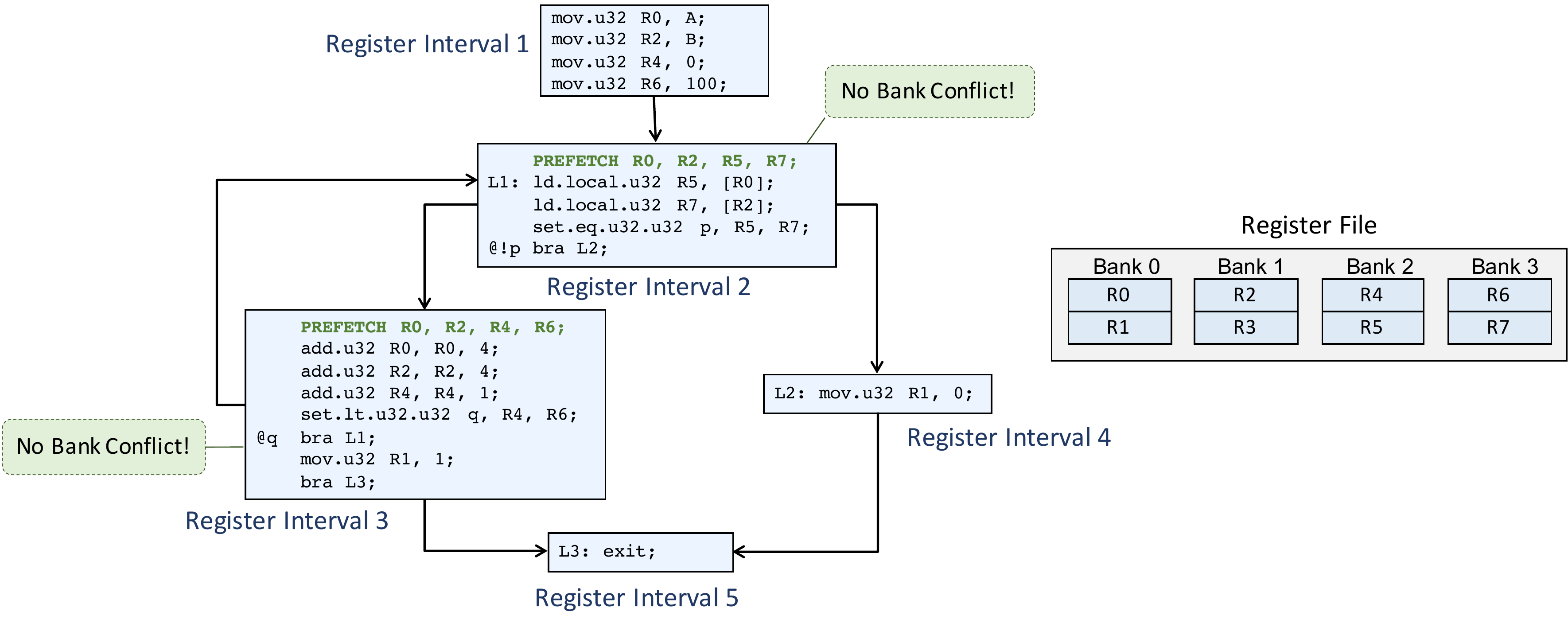}
	\caption{Register-interval CGF of PTX code in Listing \ref{code} after register renumbering.}
	\label{fig:RegisterIntervals-after}
\end{figure}

Figure \ref{fig:ICG} (bottom left side) shows the ICG of the example code in Listing~\ref{code}. The ICG shows that, for example, \texttt{R0} and \texttt{R1} cannot reside in the same register bank as \texttt{R0} and \texttt{R1} are both live at the beginning of register-interval \#2.  We color the ICG with $N_B$ colors, i.e., four colors in the example.  
We consider green, blue, yellow, and red colors for the main register banks \#0, \#1, \#2, and \#3, respectively (See Figure~\ref{fig:ICG} on top). Figure \ref{fig:ICG} (middle) shows the colored ICG specifying which bank each register should reside in to minimize register bank conflicts. We renumber each register according to its ICG color. Figure~\ref{fig:ICG} (right side) shows the ICG after renumbering. For example, \texttt{R1} is  renumbered to \texttt{R2} to resolve its bank conflict with \texttt{R0}.  Figure~\ref{fig:RegisterIntervals-after} shows the CFG of program after register renumbering. In the new CFG, all register bank conflicts existing in register-interval \#2 and register-interval \#3 are completely resolved.

\newpage

\begin{figure*}[b]
\centering
\includegraphics[trim=5mm 15mm 7mm 5mm,width=\linewidth]{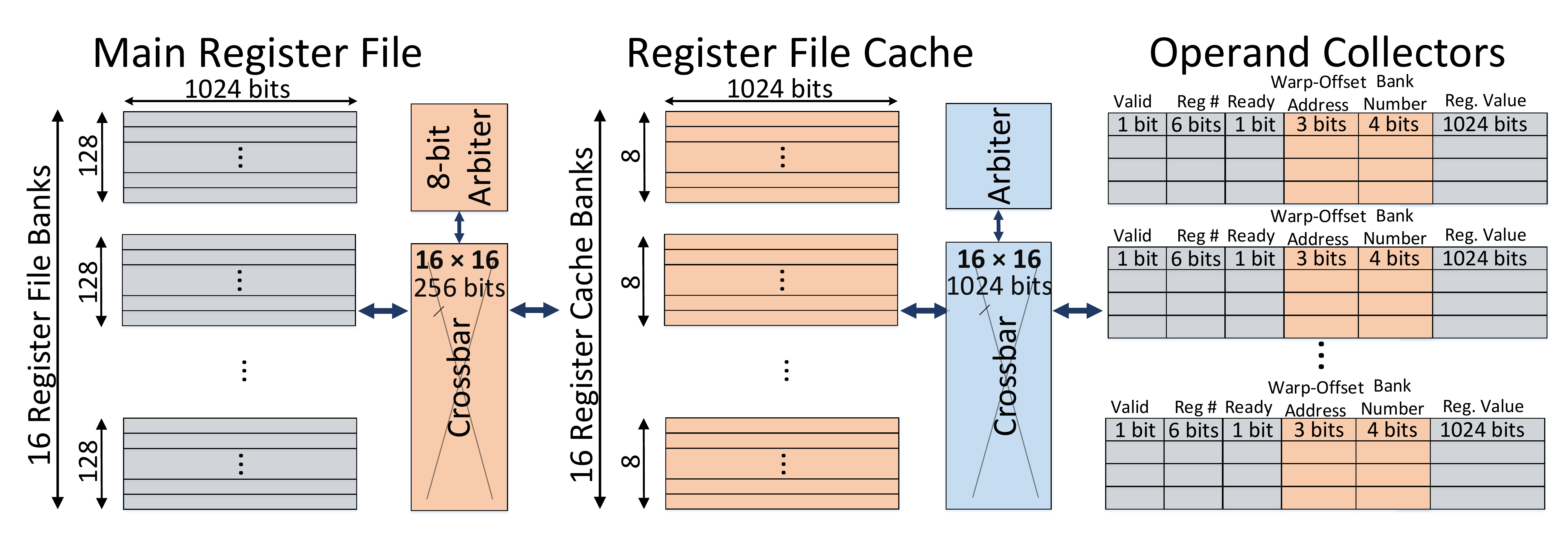}
\caption{\titleShort{} architecture. Figure assumes 8 active warps, 256 architectural registers per warp, 16 register file (cache) banks, and 16 operand collectors.}
\label{fig:RFC}
\end{figure*}

\section{Hardware Implementation}
\label{sec:Imp}
In this section, we discuss the hardware implementation of \titleShort{} in detail.

\subsection{Register File Microarchitecture}
\label{sec:ltrf-arch}
\begin{figure}[h]
\includegraphics[trim=5mm 10mm 7mm 5mm,width=0.7\linewidth]{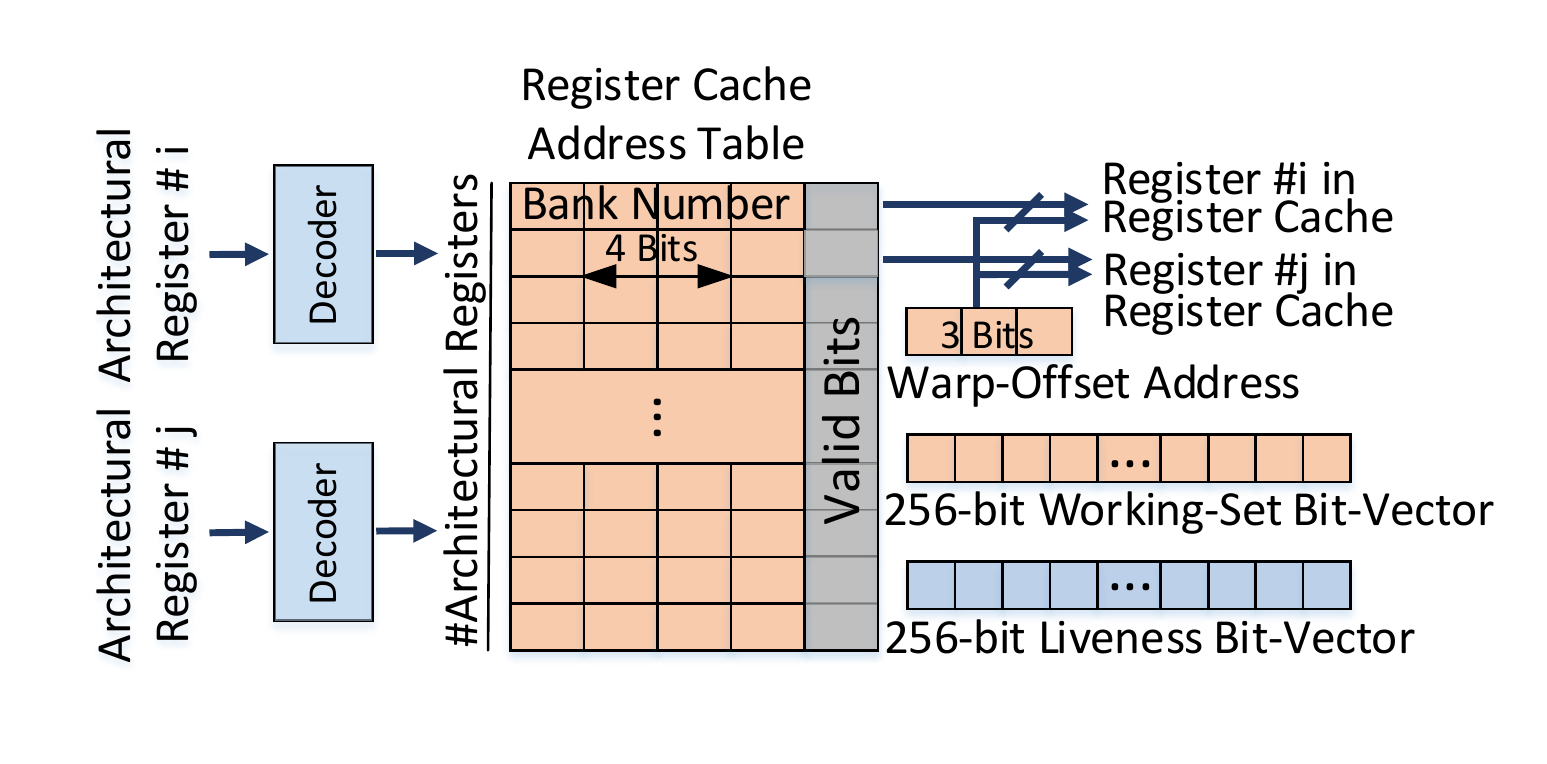}
\caption{Warp control block.}
\label{fig:MapTable}
\end{figure}

\textbf{Register File Cache.} Figure~\ref{fig:RFC} illustrates an example \titleShort{} architecture. We show our added components to the baseline register file architecture in orange color. The register file cache is composed of $\#Registers\_per\_Interval$ banks (e.g., 16 banks in the figure) where each bank hosts $\#Active\_Warps$ registers (e.g., 8 1024-bit registers in the figure). \titleShort{} interleaves registers belonging to a single warp across the cache banks, and hence, each register bank houses no more than one register of a warp. Register file cache banks are connected to the operand collectors via a crossbar.\\ 
\textbf{Warp Control Block.} A key structure in \titleShort{} design is the \textit{Warp Control Block (WCB)}, shown in Figure~\ref{fig:MapTable}. The purpose of the WCB is to maintain metadata for each warp required for controlling the register prefetching process and finding the position of the architectural registers in the register cache. To this end, WCB is composed of the \textit{register cache address table}, the \textit{working-set bit-vector}, and the \textit{liveness bit-vector}. The register cache address table is a 256-entry table per warp that keeps the register file cache bank number for each warp's architectural registers. The register cache address table has as many entries as the maximum number of architectural registers allocated to a warp. All cached registers of a warp have the same offset in all register file cache banks. Thus, for each register, the table only needs to keep track of the $\ceil{\log_2 \#Registers\_per\_Interval}$-bit (e.g., 4-bit in Figure~\ref{fig:MapTable}) index of the register file cache \emph{bank number} where that register is located. WCB also contains one $\ceil{\log_2 \#Active\_Warps}$-bit (e.g., 3-bit in Figure~\ref{fig:MapTable}) entry to track the offset of that warp's registers inside the banks (called \emph{warp-offset address}). The \emph{working-set bit-vector} holds a valid bit for each register to indicate whether it has already been prefetched during the prefetch phase. Since most of the instructions have two read operands, we provide two read ports for each  register cache address table. Any instruction that operates on more than two operands must fetch the register file cache addresses of all operands over multiple cycles.

In \titleShort{}+, which considers the liveness of operands, each warp maintains a \emph{liveness bit-vector} that keeps track of the liveness of each architectural register in the WCB, as depicted in Figure~\ref{fig:MapTable}. This vector is initially cleared (i.e., all registers are marked as dead) when the warp starts execution, and it is updated as the warp executes (\S~\ref{sec:PR}).\\
\textbf{Operand Collector Modifications.}
We augment each operand collector (Figure~\ref{fig:RFC}, right) with 
 $\ceil{\log_2 \#Registers\_per\_Interval}$-bit (e.g., 4 bits in the figure) \emph{bank number} and $\ceil{\log_2 \#Active\_Warps}$-bit (e.g., 3 bits in the figure) \emph{warp-offset address} to determine the location of each architectural register in the register file cache. \\
\textbf{Register File Cache Access.} As multiple warps may still try to access the same bank at any given cycle, we use an arbiter, as in conventional GPU register files, to arbitrate between accesses to register file cache banks and to resolve bank conflicts.
When an operand collector is allocated to a warp, it probes the register cache address table in the corresponding WCB to get the locations of registers inside the register cache. After reading the registers' locations, the operand collector participates in the arbitration phase to 1) resolve bank conflicts and 2) access the register file cache to read the operands.

\subsection{Software-Triggered Prefetch Mechanism}
\begin{figure}[t!]
\includegraphics[trim=5mm 5mm 7mm 5mm,width=0.7\linewidth]{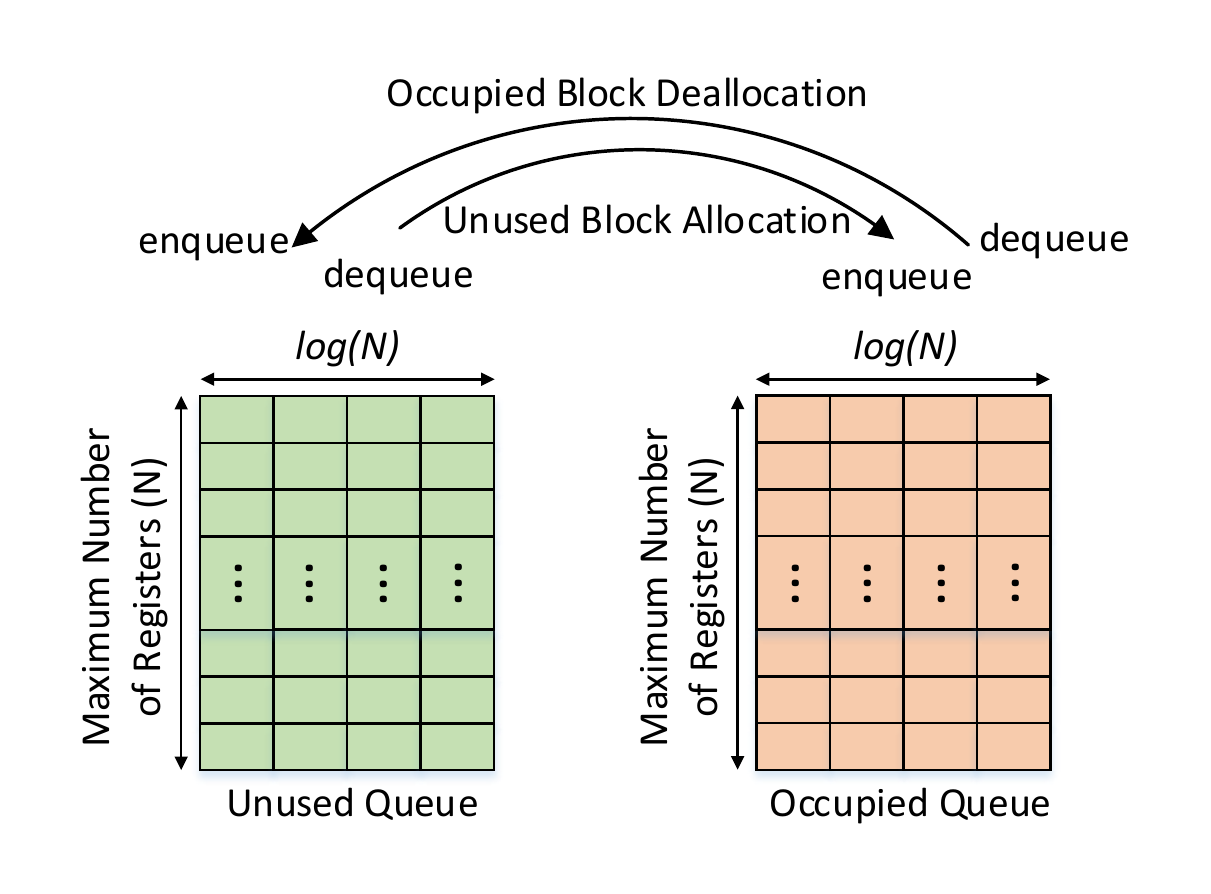}
\caption{Address allocation unit.}
\label{fig:AAU}
\end{figure}

\textbf{Executing Prefetch Operations.} When a warp reaches a prefetch operation, the GPU must load the warp's registers into the register file cache as indicated by the prefetch bit-vector. Initially, the prefetch bit-vector is decoded into a list of indices (IDs) of registers that need to be loaded. Once the register indices are identified, they must be allocated space in the register file cache, and the warp's register cache address table in the WCB must be properly filled. After allocating register file cache space, the registers can be read from the main register file to fill the register cache. When a register is prefetched completely, the corresponding valid bit in the WCB is set. After all registers indicated by the prefetch bit-vector are prefetched, the warp becomes ready to execute, and all subsequent register accesses of that warp are served from the register file cache. In \titleShort{}+, whenever a warp performs a prefetch operation, it queries the liveness bit-vector and prefetches \emph{only} the registers that are marked as live. For dead registers, it is sufficient to allocate the register file cache space, without fetching data. \\
\textbf{Register File Cache Space Allocation.} Every cached register in a register-interval must be assigned a place in the register file cache. In our design, this mechanism is equivalent to allocating one register file cache bank for each cached register as we interleave the registers of a single warp across banks to minimize register file cache bank conflicts. We employ the \emph{Address Allocation Unit}, depicted in Figure~\ref{fig:AAU}, for \emph{each warp} to implement this mechanism. The Address Allocation Unit is composed of two queues: the \emph{unused} queue keeps track of free banks, while the \emph{occupied} queue keeps track of allocated banks. Initially, the unused queue is full, and the occupied queue is empty. On an allocation, we allocate the head of the unused queue to the new register and move that entry to the occupied queue. On a deallocation, we move the deallocated register entry back to the unused queue. The same mechanism is used to allocate warp-offset addresses to warps. There, we use a \emph{global} Address Allocation Unit that is shared by all warps.

\noindent\textbf{Interconnect.} We use an $\#Active\_Warps$-bit arbiter (e.g., 8-bit arbiter in Figure~\ref{fig:RFC}, left) to arbitrate among warps to fill the register file cache. Registers are loaded into the register file cache from the main register file via a crossbar network. In order to design this crossbar, we calculate the number of accesses to the main register file in \titleShort{} and the baseline architecture. Our experimental results show that \titleShort{} reduces the number of accesses to the main register file by 4$\times$-6$\times$ (as most of the accesses are serviced through the register file cache) and the bandwidth of the 1024-bit crossbar in the baseline architecture \emph{without} register caching~\cite{gpuwattch,GPGPU-Sim} is utilized by up to 85\%. As a result, we can reduce the bandwidth of the main register file crossbar by 4$\times$ without hurting performance. On the downside, the narrower crossbar would exhibit a traversal latency 4$\times$ larger (from 1 cycle to 4 cycles) than a wide crossbar in typical scenarios, and far larger latency when the crossbar is saturated and queuing effects become dominant. However, due to the latency-tolerant nature of our design and the fact that register file access latencies are dominated by bank access times rather than crossbar traversals, we find that the longer traversal latency of a narrower crossbar is \emph{not} a significant performance issue. Because warp registers are interleaved across the register cache banks, \titleShort{} improves the \emph{parallelism of accesses in the interconnect}. \\
\textbf{Warp Stall.} A warp that is stalled becomes inactive and loses its slots in the register file cache. In that case, it must perform three steps. First, it writes back its \emph{live} registers to the main register file. Second, it releases its register file cache slots. Third, it clears all the valid bits in the register cache address table of the WCB. Whenever a warp goes from being inactive to active, and it finds itself in the middle of a register-interval, it must refetch all its specified registers in its working-set bit-vector that are still live from the main register file. This is done using the warp's working-set and liveness bit-vectors in WCB.

\subsection{Overheads}
\label{sec:imp-overhead}
\textbf{Code Size.} \titleShort{} increases the average code size by 7\% if only prefetch bit-vectors are inserted into the code and 9\% if the bit-vectors are accompanied by prefetch instructions. Note that, in order to be able to only insert the bit-vectors into the code, the ISA has to be redesigned and an extra bit has to be embedded into \emph{all} instructions, as explained in \S\ref{sec:PR}. To measure the effect of the increased code size on the GPU performance, we execute the original and the modified programs on the baseline architecture using GPGPU-Sim~\cite{GPGPU-Sim}. Our experimental results show that the larger code size results in 0.2\% average (up to 1\%) performance degradation, which is negligible. We do not add a new instruction to pass liveness information to hardware. We use the same mechanism as the prior work~\cite{cache1}, which embed liveness information in the instruction using free spaces in the instruction code. As a result, marking dead registers does not increase the code footprint and the number of executed instructions.\\
\textbf{Storage Cost.} \titleShort{} requires a WCB for every warp, shown in Figure~\ref{fig:MapTable}. Each WCB contains one 5-bit entry per architectural register, 3-bit for the warp-offset address, and working-set and liveness bit-vectors, each with one bit per register. The total storage overhead of the WCB for each SM in an example modern architecture, which supports 64 warps with 256 registers per warp, is 114880 bits ($64\times(256\times5+3+256+256)$), around 5\% of the area consumed by the 256KB baseline register file. \\
\textbf{Latency Overhead.} According to our analysis with CACTI~\cite{muralimanohar2009cacti}, the WCB can be accessed within one extra clock cycle. Hence, it adds negligible performance overhead in accessing the registers.\\
 \textbf{Area/Power Cost.}
In order to measure the area and power overheads of \titleShort{}, we functionally model all the added components (i.e., WCB, the additional crossbar, address allocation units, the arbiter, additional entries in the operand collectors, and register file cache) in GPU-Wattch~\cite{gpuwattch}. In total, \titleShort{} occupies 16\% more area than our baseline GPU register file (i.e., Configuration \#1 in Table~\ref{tab:RF_archs}) using the same main register file size and technology. In terms of power consumption, despite the added structures, \titleShort{} consumes 23\% less power compared to the baseline register file. \titleShort{}'s improvement in power consumption is due to reducing the number of accesses to the main register file by 4$\times$-6$\times$.

\section{Methodology}
\label{sec:method}
\noindent\textbf{Simulation.} We evaluate our techniques using the GPGPU-Sim V3.2.2~\cite{GPGPU-Sim} cycle-level simulator for GPUs. Table \ref{tab:param} provides the details of our baseline GPU configuration. We model our baseline after an NVIDIA Maxwell-like architecture~\cite{Nvidia_white_paper_maxwell}. 
We modify the microarchitecture of the conventional register file in GPGPU-Sim to implement the \titleShort{} microarchitecture depicted in Figure~\ref{fig:RFC}.

\begin{table}[!h]
\renewcommand{\arraystretch}{1}
\small
\centering
\caption{Simulated system configuration.}
\vspace{-6pt}
\begin{tabular}{|*{2}{c|}}
\hline
Number of SMs&24\\ \hline
Core clock&1137 MHz\\ \hline
Scheduler&Two-level~\cite{cache1,narasiman2011improving}\\ \hline
Number of warps per SM&64\\ \hline
Register file size&256KB per SM (65536 registers)\\ \hline
Register file cache size&16KB per SM (4096 registers)\\ \hline
Shared memory size&64KB per SM\\ \hline
L1D Cache&4-way, 16KB, 128B line\\ \hline
L1I Cache&4-way, 2KB, 128B line\\ \hline
LLC&8-way, 2MB, 128B line\\ \hline
{Memory Model}     & 8 GDDR5 MCs, \\ & FR-FCFS~\cite{fr-fcfs1,fr-fcfs2}, 2700 MHz \\ \hline
{GDDR5 Timing} & $t_{CL}$=12, $t_{RP}$=12, $t_{RC}$=40,           \\
(in nanoseconds)                 & $t_{RAS}$=28, $t_{RCD}$=12, $t_{RRD}$=6    \\ \hline
Number of active warps & 8 per SM \\ \hline
Number of registers & 16 \\
in a register-interval &  \\ \hline
\end{tabular}
\label{tab:param}
\end{table}

We use the compiler in GPGPU-Sim to implement our software prefetching mechanism. To this end, we process the CFG of the register-allocated PTX code to form the register-intervals\footnote{We open source the C implementation of register-interval creation in~\cite{register-interval-github}.} and insert prefetch bit-vectors at the start of each register-interval. \\
\textbf{Benchmarks.} We run 35 benchmarks from CUDA SDK \cite{GPGPU-Sim}, Rodinia \cite{rodinia}, and Parboil \cite{parboil} benchmark suites and classify them into two groups, \emph{register-sensitive} and \emph{register-insensitive}, based on whether or not the register file limits the achievable TLP. Enlarging GPU register file improves the thread-level parallelism of register-sensitive applications. However, not all register-sensitive applications benefit the same from higher thread-level parallelism in terms of GPU performance (see \S~\ref{sec:main-perf} for more detail). We randomly select nine workloads from the register-sensitive group, and five workloads from the register-insensitive one. 
\\
\textbf{Comparison Points.} We evaluate (1)~a baseline (BL) architecture that models a GPU with a conventional non-cached register file. To provide a fair comparison of this baseline to other register file cache based designs, we add the amount of space dedicated for the register file cache in \titleShort{} (16KB) to the main register file capacity in the BL architecture, (2)~a design with a 16KB register file cache (RFC) \emph{without} any prefetching mechanisms, similar to the architecture proposed in~\cite{cache1}, (3)~\titleShort{} with a 16KB register file cache. (4)~LTRF$_{conf}$, an enhanced version of \titleShort{} in which, we reduce the register bank conflicts in performing prefetch operations using a compiler pass after register-interval creation pass (\S~\ref{sec:BC}). (5)~a GPU with an \emph{Ideal} register file architecture that allows us to increase the register file capacity to any size (i.e., 8$\times$ in our evaluations) with \emph{no latency overhead}. Please note that all these designs have 24 SMs and 64 warps per SM.\\
\textbf{Design Points.} To show the benefits of \titleShort{}, we select different memory technologies (i.e., TFET~\cite{TFET} and DWM~\cite{DWM}) as use-cases. The reason for selecting different memory technologies is that they let us evaluate a wider range of access latencies; we believe that using different memory technologies is one of the concrete potential futuristic use-cases since these technologies are finding their ways into commercial products due to limitations of existing technologies in scaling to smaller dimensions~\cite{lee2009architecting,lee2010phase,qureshi2009scalable,kultursay2013evaluating,tavakkol2018enabling}. Yet, our optimizations do not depend on the existence of a new memory technology. We can have large register files where \titleShort{} would be useful even in conventional technologies by paving the way for different register file optimization techniques, such as register file compression~\cite{RF_comp,CABA,pekhimenko2016case,pekhimenko2015toggle}, register file virtualization~\cite{Zorua,virtual_RF1,khorasani2018regmutex,regless}, and register file power-gating~\cite{gate3,gate1,10.1145/3291606,kayiran2016muc,7092631,7273522}. We increase the register file size from 256KB to 2MB by using the register file configurations \#6 and \#7 from Table~\ref{tab:RF_archs}. Configuration \#6 allows us to increase the register file size by 8$\times$ while keeping the power consumption almost unchanged. Configuration \#7, on the other hand, results in less power/area consumption compared to the baseline SRAM-based 256KB register file. We use these design points as realistic baselines for our performance analysis. 
We carefully model the register prefetching latency. Three important factors that affect prefetching operation delay are (1) the register bank access latency, (2) the register bank conflict, and (3) the transferring latency. We model the register bank access latency of each memory technology using CACTI~\cite{muralimanohar2009cacti} and NVSim~\cite{dong2014nvsim}. We then import the calculated numbers to the GPGPU-Sim~\cite{GPGPU-Sim} to consider the effect of register bank conflicts, as well. Regarding the transferring latency, we use the booksim simulator~\cite{booksim} embedded in the GPGPU-Sim.\\
\textbf{Performance Metrics.} We use IPC as the performance metric to evaluate different register file designs. We evaluate our compiler algorithms by measuring the size of the generated register-intervals.

\section{Evaluation}
\label{sec:eval}
We present the effectiveness of five different mechanisms: BL, RFC, \titleShort{}, \titleShortnew{}, and Ideal. \S~\ref{sec:main-perf} shows the overall effect of \titleShort{}~on GPU performance.  
\S~\ref{subsec:ltrf-perf} analyzes the effectiveness of \titleShort{} at tolerating the latency of the main register file. \S~\ref{sec:RB-eval} evaluates the number of register blocking events in \titleShort{} and \titleShortnew{}. \S~\ref{sec:sensitivity} provides sensitivity analysis on the size of the register file cache. \S~\ref{sec:interval} analyzes the number of instructions in register-intervals.  
\S~\ref{sec:ltrf-vs-software} provides a comparison between \titleShort{} and other software-managed register caching schemes.

\begin{figure*}[t]
\begin{tabular}{c}
\includegraphics[trim=30mm 116mm 14mm 80mm,width=\linewidth]{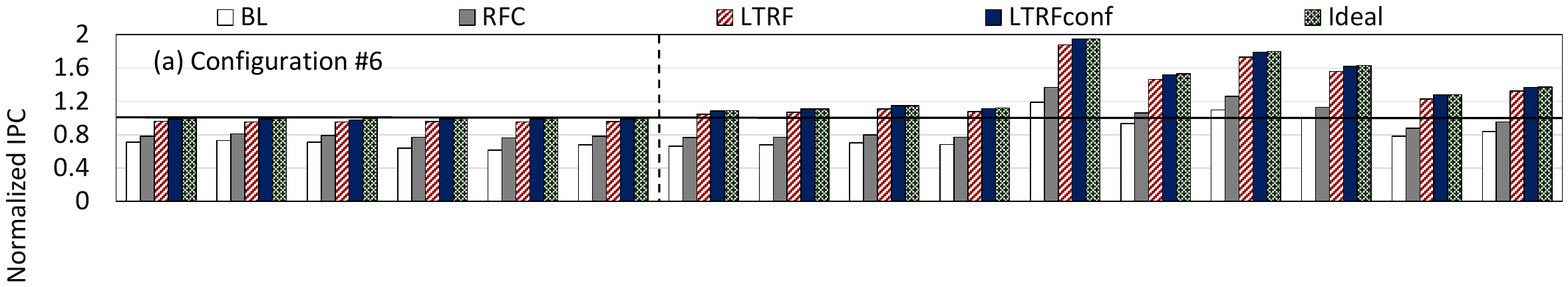}
\\
\includegraphics[trim=30mm 81mm 14mm 55mm,width=\linewidth]{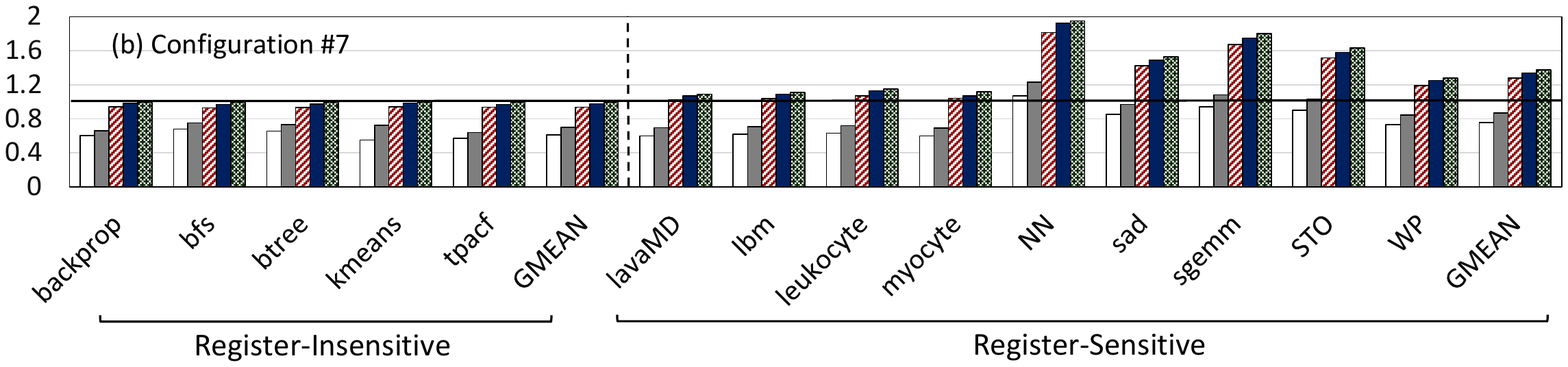}
\\
\end{tabular}
\caption{IPC of BL, RFC, \titleShort{}, \titleShortnew{}, and Ideal using the main register file configurations \#6 and \#7 from Table~\ref{tab:RF_archs}, normalized to the baseline architecture of configuration \#1 with 16KB additional register file capacity.}
\label{fig:moreIPC}
\end{figure*}

\subsection{Overall Effect on GPU Performance}
\label{sec:main-perf}
To evaluate the effect of larger register files on GPU performance, we increase the register file size from 256KB to 2MB by using the register file configurations \#6 and \#7 from Table~\ref{tab:RF_archs}. Figure~\ref{fig:moreIPC} compares the normalized IPC of BL, RFC, \titleShort{}, and \titleShortnew{} designs when used on top of these two configurations. In this figure, we normalize the IPC results to the IPC results of the baseline architecture of configuration \#1 in Table~\ref{tab:RF_archs}, without any register caching, with one modification: we add the register file cache capacity (i.e., 16KB) used in the other mechanisms to the 256KB register file size of configuration \#1. Ideal bars show the IPC of an idealized version of configuration \#1 with 8$\times$ the register file capacity but \emph{no} increase in latency (i.e., access latency remains constant after increasing register file size by 8$\times$). We make three major observations. First, \titleShort{} provides almost the same IPC performance as the Ideal design when we employ configuration \#6. \titleShort{} improves IPC by 32\%, on average. The IPC improvement of \titleShort{} is due to two main reasons. (1) The larger register file enables both more registers per thread and more warps executing in parallel. (2) \titleShort{} effectively tolerates the higher access latency of configuration \#6. Second, for the register-insensitive workloads that do \emph{not} benefit from a larger register file (e.g., \texttt{btree} and \texttt{kmeans}), the performance overhead of increasing the register file size is minimal if we use \titleShort{} and \titleShortnew{} as opposed to RFC. Third, \titleShortnew{} improves the performance of \titleShort{} by an average 3.8\% and 4.8\% for configurations \#6 and \#7, respectively. These results clearly show the positive effect of reducing the number of register blocking events on \titleShort{} performance. Fourth, \titleShort{} and \titleShortnew{} effectively enable the use of configuration \#7, which reduces the register file area by 75\%.  
For this configuration, \titleShort{} and \titleShortnew{} improve performance by 28\% and 34\% over the baseline, on average, respectively. We conclude that \titleShort{} enables a high-capacity and high-latency main register file while providing high performance.

\subsection{Effect of \titleShort{} on Register File Access Latency}
\label{subsec:ltrf-perf}

\begin{figure*}[h]
\centering
\includegraphics[trim=37mm 88mm 24mm 82mm,width=\linewidth]{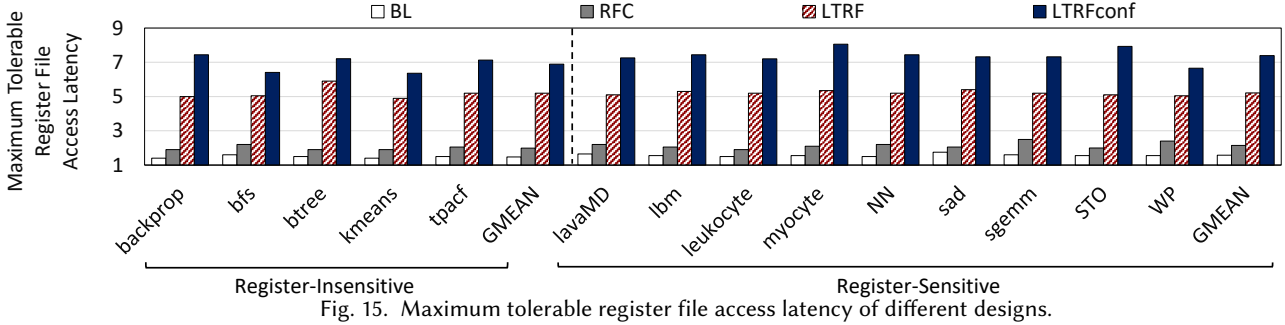}
\caption{Maximum tolerable register file access latency of different designs.}
\label{fig:IPCs-MR}
\end{figure*}

To show the effectiveness of \titleShort{} at tolerating register file access latency, we define a new metric: the \emph{maximum tolerable register file access latency}. This is the relative latency\footnote{Relative to the baseline of configuration \#1 with 16KB additional register file capacity} of the main register file that leads to at most 5\% performance (IPC) loss for each workload we examine. Note that this metric is different for each design, depending on the latency tolerance of the design. We increase the main register file access latency while keeping the main register file size constant. 
Figure~\ref{fig:IPCs-MR} compares the maximum tolerable register file access latency of different designs for various benchmarks.

We make three major observations. First, the maximum tolerable register file access latency for \titleShort{} is 5.3$\times$, on average. This result indicates that \titleShort{} can 1) effectively bring the registers to the register file cache before they are accessed and 2) hide the latency of the register access to the main register file by executing other active warps. Second, the maximum tolerable register file access latency for \titleShortnew{}, in which, we minimize register bank conflicts, is 6.9$\times$, on average, indicating that resolving register bank conflicts can effectively improve the opportunity of \titleShort{} in tolerating the register file access latency.
Third, the maximum tolerable register file access latency for RFC is 2.1$\times$, on average, which shows that the register file cache hit rate in the RFC design is \emph{not} large enough to hide main register file access latencies that are greater than 2.1$\times$.

We conclude that \titleShort{} and \titleShortnew{} are able to tolerate long main register file access latencies. Thus, they can enable aggressive optimizations that increase register file capacity in exchange for higher access latency.

\subsection{Register Bank Conflicts in \titleShort{} vs. \titleShortnew{}}
\label{sec:RB-eval}
\begin{figure}[h]
\begin{tabular}{c c}
\includegraphics[trim=25mm 50mm 25mm 50mm, width=0.5\linewidth]{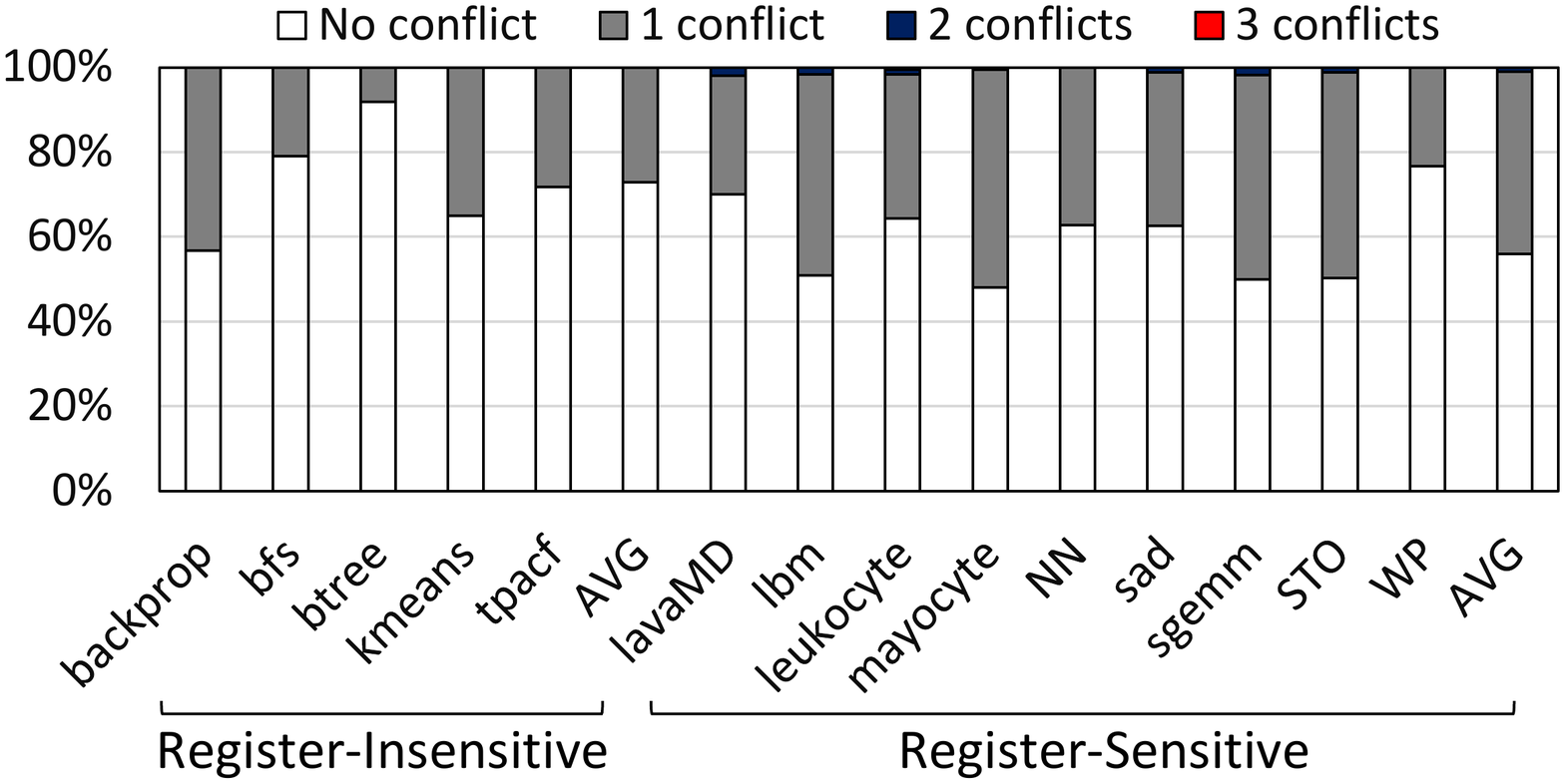}&
\includegraphics[trim=25mm 50mm 25mm 50mm, width=0.5\linewidth]{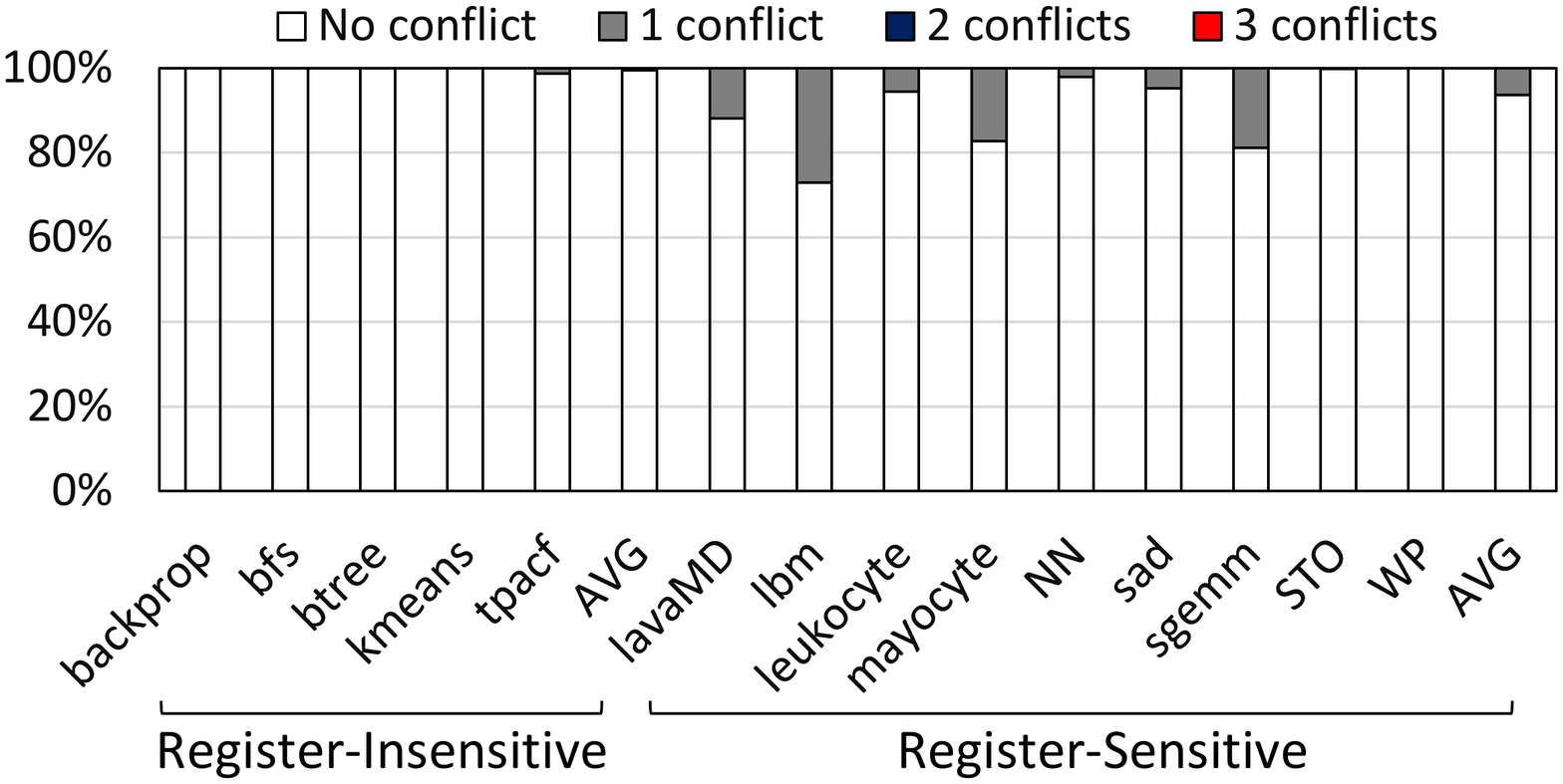}\\
\begin{footnotesize}
(a)
\end{footnotesize}&
\begin{footnotesize}
(b)
\end{footnotesize}\\
\includegraphics[trim=25mm 50mm 25mm 50mm, width=0.5\linewidth]{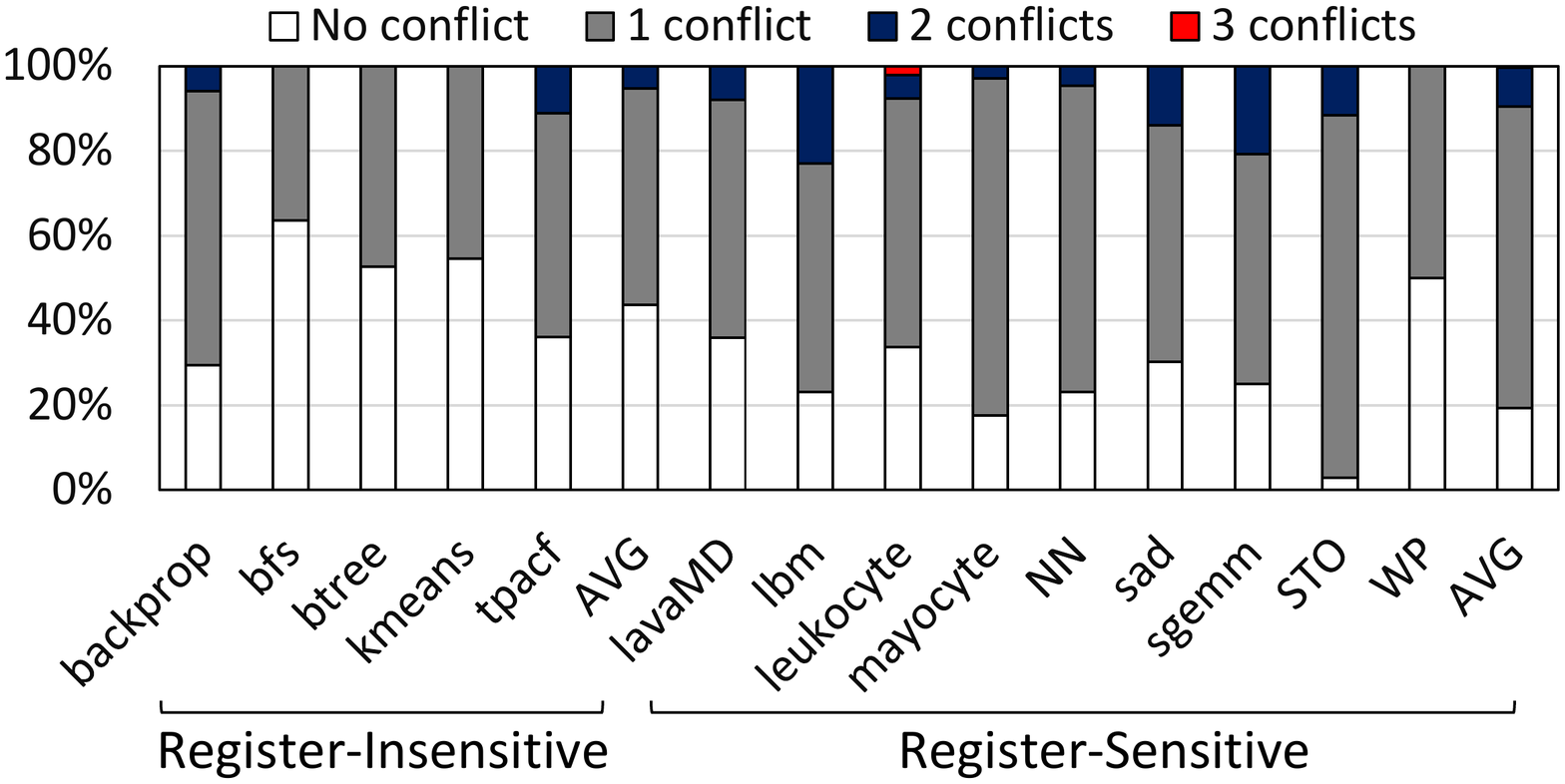}&
\includegraphics[trim=25mm 50mm 25mm 50mm, width=0.5\linewidth]{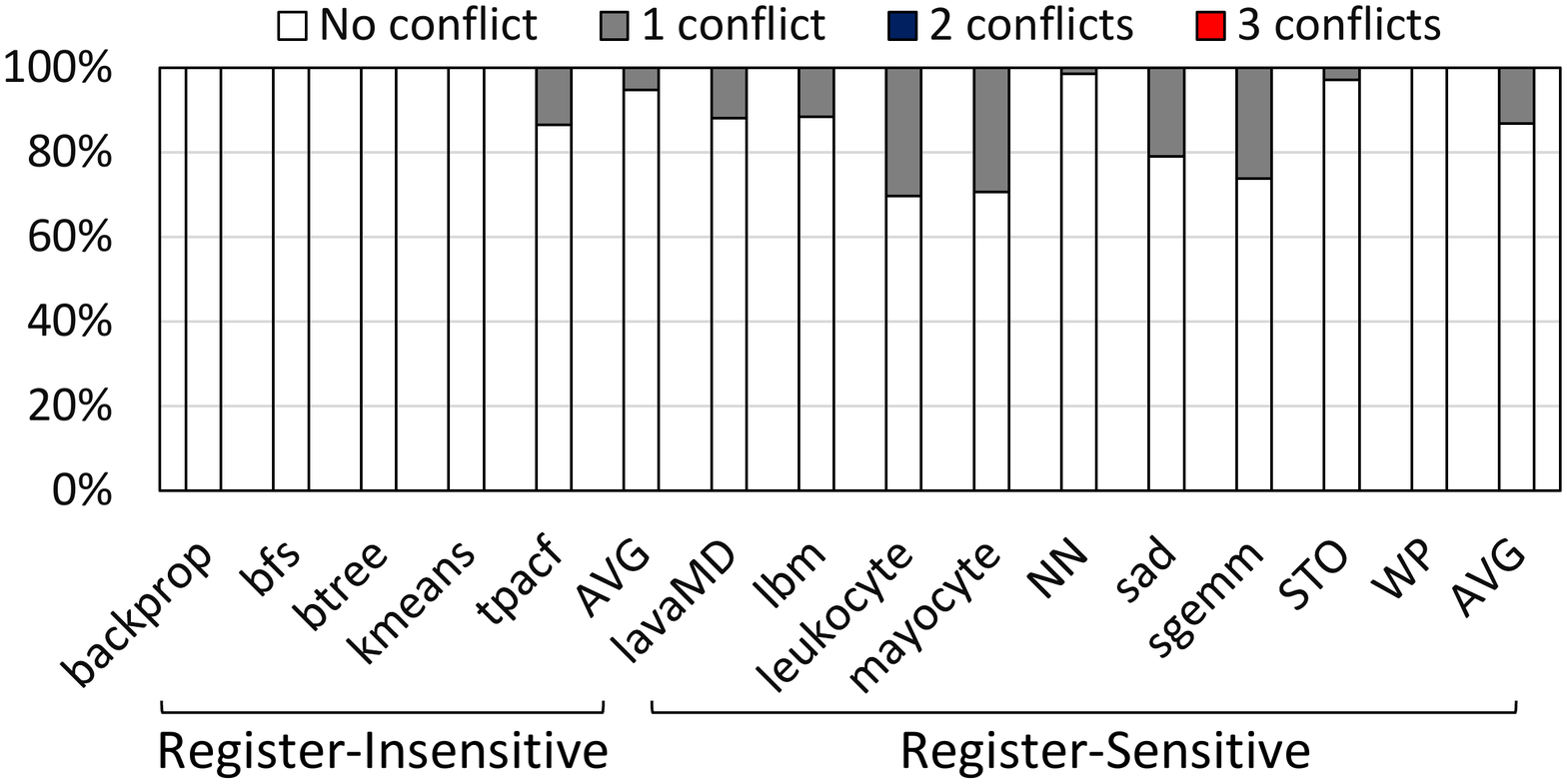}\\
\begin{footnotesize}
(c)
\end{footnotesize}&
\begin{footnotesize}
(d)
\end{footnotesize}\\
\includegraphics[trim=25mm 50mm 25mm 50mm, width=0.5\linewidth]{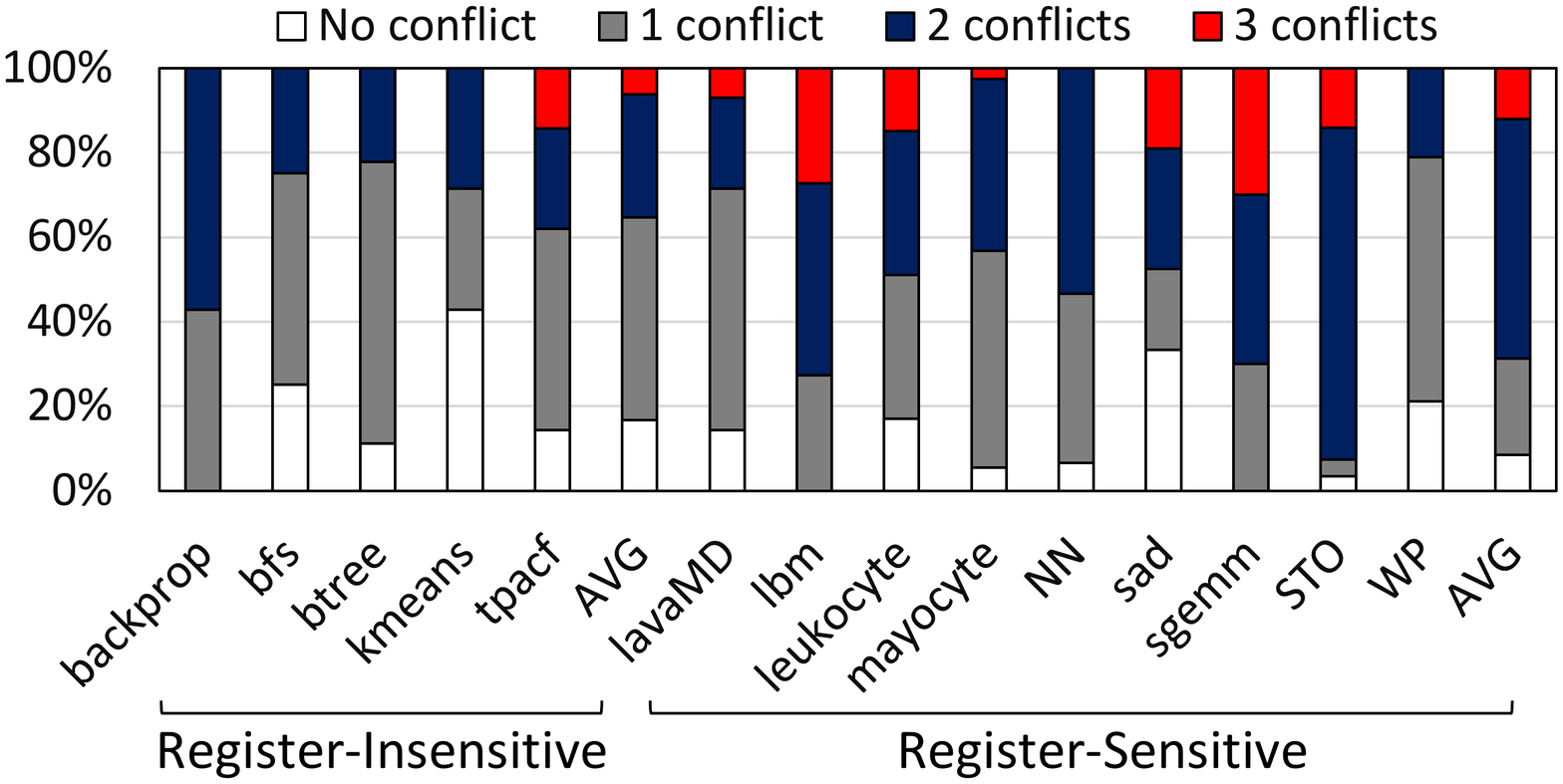}&
\includegraphics[trim=25mm 50mm 25mm 50mm, width=0.5\linewidth]{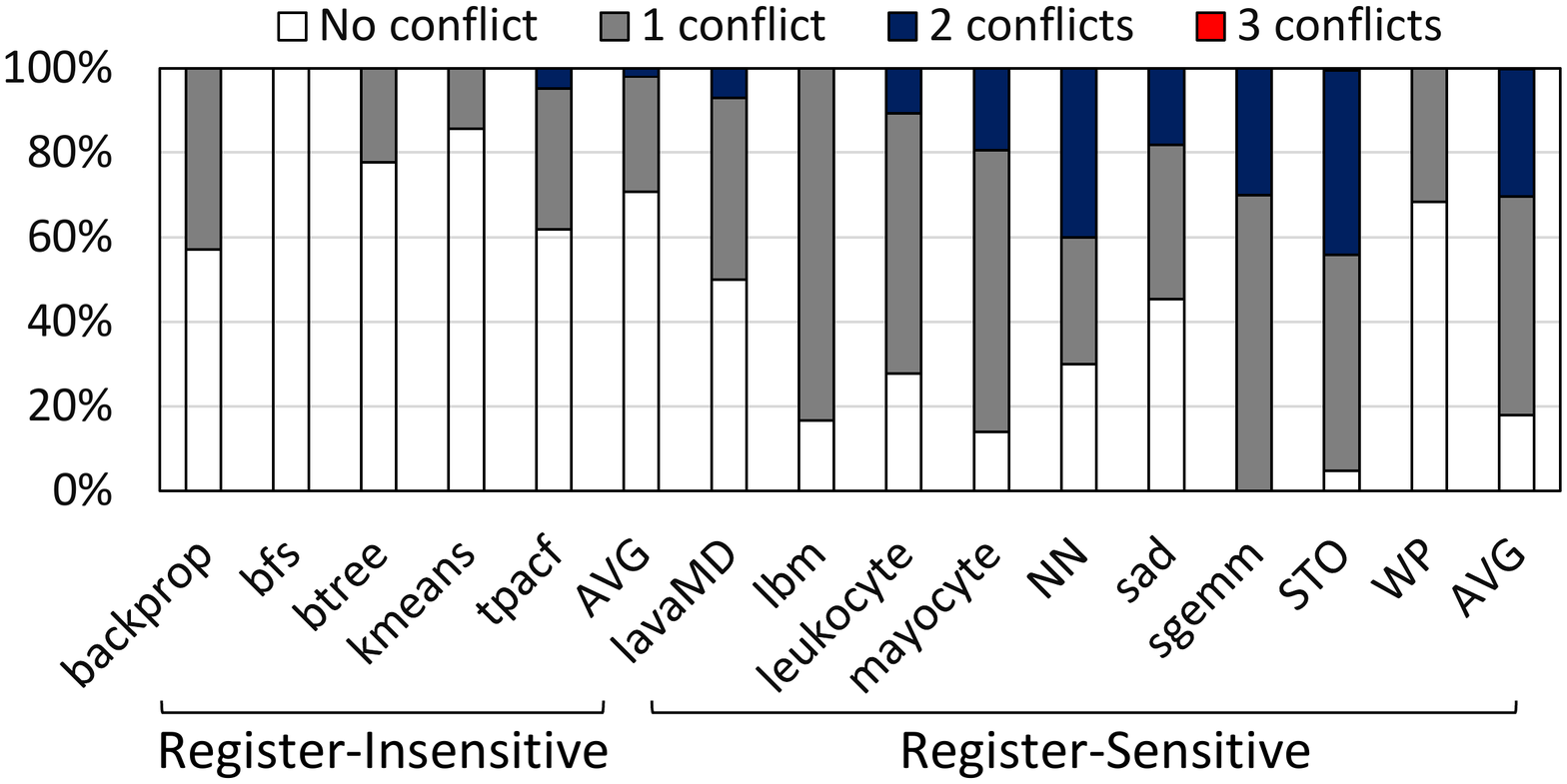}\\
\begin{footnotesize}
(e)
\end{footnotesize}&
\begin{footnotesize}
(f)
\end{footnotesize}\\
\end{tabular}
\caption{Distribution of the number of register bank conflicts in register-intervals for different workloads; (a) \titleShort{} with 8 registers allowed in each register-interval, (b) \titleShortnew{} with 8 registers allowed in each register-interval, (c) \titleShort{} with 16 registers allowed in each register-interval, (d) \titleShortnew{} with 16 registers allowed in each register-interval, (e) \titleShort{} with 32 registers allowed in each register-interval, and (f) \titleShortnew{} with 32 registers allowed in each register-interval.}
\label{fig:BC-designs}
\end{figure}

To evaluate the effect of \titleShortnew{} on resolving register bank conflicts, we compare the number of register bank conflicts for different workloads using \titleShort{} and \titleShortnew{}. We perform the experiments using various numbers of registers allowed in each register-interval, i.e., 8, 16, and 32, with a fixed number of register banks (i.e., 16 banks). Figure \ref{fig:BC-designs}(a-f) illustrates the results.

We make two key observations. First, on average, 58\%, 23\%, and 9.4\% of prefetch operations experience conflict-free register bank access in \titleShort{} using 8, 16, and 32 registers allowed in each register-interval, respectively. Our compile-time register renumbering technique enables an average of 95\%, 88\%, 24\% of prefetch operations to have conflict-free register bank access when 8, 16, and 32 registers allowed in each register-interval, respectively. Second, our register renumbering technique reduces the maximum number of register bank conflicts in a prefetch operation from 3, 3, and 3 to 1, 1, and 2 for 8, 16, and 32 registers allowed in each register-interval, respectively. We conclude that our compile-time register renumbering technique embedded in \titleShortnew{} is very effective at resolving register bank conflicts, and paves the way for increasing the number of registers allowed in each register-interval.

\subsection{Sensitivity to Register File Cache Size}
\label{sec:sensitivity}
\begin{figure}[h!]
\centering
\includegraphics[trim=21mm 62mm 28mm 57mm,width=0.7\linewidth]{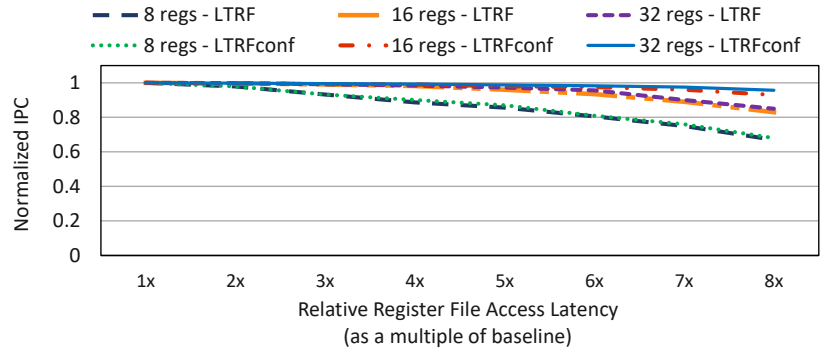}
\caption{Normalized IPC using \titleShort{} with various main register file access latencies and number of allowed registers in each register-interval.}
\label{fig:sen_regs}
\end{figure}

We explore the effect of the register file cache size on performance in two ways: (1) varying the number of registers allowed in each register-interval (default is 16), (2) varying the number of active warps with allocated storage space in the register cache.
Figure~\ref{fig:sen_regs} reports the average IPC when we vary the number of registers allowed in each register-interval for \titleShort{} and \titleShortnew{}. We make four observations. First, when the number of registers allowed in each register-interval is 8, the effectiveness of \titleShort{} degrades significantly, as the main register file access latency increases. This is mainly because a small number of registers results in a small register-interval size. Hence, prefetch operations become more frequent, and hiding their latency becomes more difficult, especially for slow main register files. Second, \titleShortnew{} and \titleShort{} work almost the same when the number of registers allowed in each register-interval is 8 as we do \emph{not} have much register bank conflict in this case. Third, increasing the number of registers allowed in each register-interval does \emph{not} necessarily translate to better performance for \titleShort{}. This is mainly because more registers result in more main register file bank conflicts during the prefetch operation, increasing prefetch latency. Therefore, larger register-interval sizes may not always be enough to hide larger prefetch latencies. Fourth, increasing the number of registers allowed in each register-interval can result in better performance of \titleShortnew{} as \titleShortnew{} attempts to reduce the number of register bank conflicts to the minimum possible value (i.e., one register bank conflict when the number of registers allowed in each register-interval is 32).

\begin{figure}[h!]
\centering
\includegraphics[trim=21mm 68mm 28mm 57mm,width=0.7\linewidth]{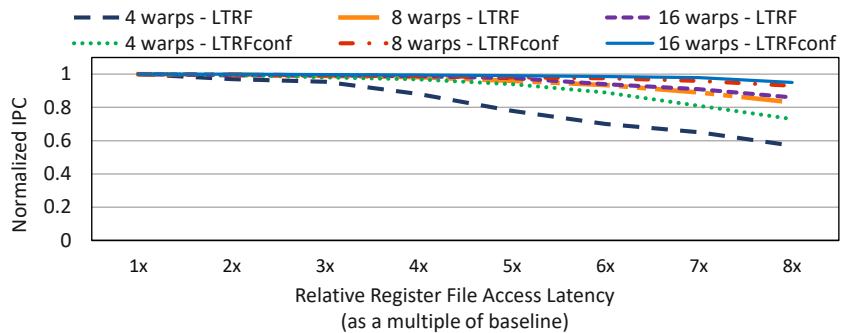}
\caption{Normalized IPC using \titleShort{} with various main register file access latencies and number of active warps.}
\label{fig:sen_warps}
\end{figure}

Figure \ref{fig:sen_warps} illustrates \titleShort{} performance sensitivity to the number of warps that have dedicated register file cache space, while keeping the dedicated space per warp constant. We make three observations. First, as the number of active warps increases from 4 to 8, IPC improves by 45.6\% and 27.4\% for the slowest main register file using \titleShort{} and \titleShortnew{}, respectively. Second, increasing the number of active warps by more than 8 does \emph{not} have a significant impact on \titleShort{} and \titleShortnew{} performance. Third, the performance improvement of \titleShortnew{} over \titleShort{} is larger for smaller number of active warps. We conclude that 8 active warps, which is the default configuration in \titleShort{}, seems enough. Hence, \titleShort{} does \emph{not} impose significant performance cost by limiting the number of active warps.

We conclude that, by making the performance impact of a slower register file more tolerable, \titleShort{} enables a large design space to architects, where tradeoffs between power, area, and latency of the register file can be explored more freely to optimize system-level goals.

\subsection{Register-Interval Length}
\label{sec:interval}
Register-intervals should be as long as possible to minimize the number of prefetch operations. We measured both the real and the optimal register-interval lengths. The \emph{real register-interval length} is the number of dynamic instructions within each register-interval. The \emph{optimal register-interval length} is the number of consecutive dynamic instructions in a kernel's execution trace that consume at most the maximum number of allowed registers in the register cache. In other words, the optimal length exposes the limitations caused by the control-flow constraints imposed on register-intervals.  Table~\ref{tab:interval-length} reports the average, minimum, and maximum lengths of the real and optimal register-intervals. We make two observations. First, the real register-interval length is 89\% of the optimal register-interval length, on average. Second, the minimum and maximum lengths of real register-intervals are 78\% and 85\% of the ones in optimal register-intervals, respectively. We conclude that the control flow constraints in creating the register-intervals do \emph{not} greatly limit the register-interval length.

\begin{table}[h!]
\small
\centering
\caption{The average, minimum, and maximum lengths of real and optimal register-intervals, in terms of dynamic instructions, for 35 workloads in CUDA SDK~\cite{GPGPU-Sim}, Rodinia~\cite{rodinia}, and Parboil~\cite{parboil} benchmark suites.}
\vspace{-6pt}
\begin{tabular}{|c || c |c |c|}
\hline 
Register-Interval Length  & Average & Minimum & Maximum \\ \hline \hline
Real & 31.2  & 7 & 45 \\ \hline
Optimal & 34.7 & 9 & 53 \\ \hline
\end{tabular}
\label{tab:interval-length}
\end{table}   

\subsection{\titleShort{} vs. SW-Managed Hierarchical Register Files}
\label{sec:ltrf-vs-software}
To distinguish the benefits of our key ideas from other software-based approaches,  
we evaluate the maximum tolerable register file access latency of two additional designs: 1) a software-managed hierarchical register file (SHRF) similar to~\cite{cache2} and 2) a version of \titleShort{} that performs prefetch operations at the end of \emph{strands}~\cite{cache2}, rather than register-intervals. SHRF~\cite{cache2} aims to reduce the number of background register swap operations between the main register file and the register file cache to provide energy efficiency and uses traditional register allocation/spilling techniques. SHRF uses strands, which are more constrained CFG subgraphs than register-intervals, since long/variable-latency operations (e.g., cache misses) and backward branches are disallowed within a strand to guarantee that the warp does \emph{not} get descheduled until the end of the strand~\cite{cache2}.

\begin{figure}[h!]
\centering
\includegraphics[trim=25mm 56mm 23mm 68mm,width=0.6\linewidth]{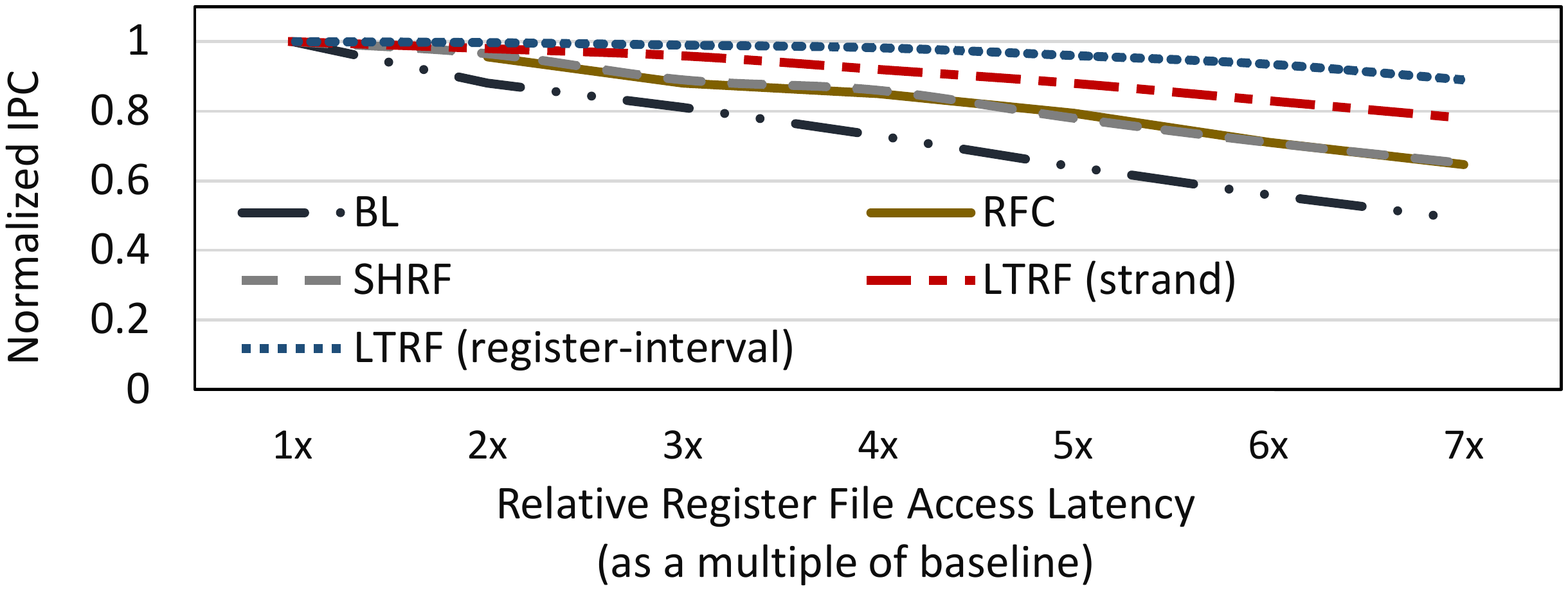}
\caption{Normalized IPC using BL, RFC, SHRF, \titleShort{} (strand), and \titleShort{} (register-interval) with various register file access latencies.}
\label{fig:LTRF_sw}
\end{figure}

Figure \ref{fig:LTRF_sw} reports the normalized IPC, averaged across our workloads (see \S~\ref{sec:method}),  
as the main register file access latency increases.

We make two observations. First, SHRF performs similarly to RFC and can tolerate latencies by up to 2$\times$ the baseline latency. Second, \titleShort{} can tolerate only 3$\times$ higher register file latency if it uses strands instead of register-intervals, as opposed to the 5.3$\times$ higher main register file latency tolerated by our \titleShort{} design that uses register-intervals.  \titleShort{} performs better using register-intervals because strands' CFG subgraphs are more constrained, and typically much smaller than the CFG subgraphs of register-intervals, increasing the number of prefetch operations, register writebacks, and register re-fetches. In particular, while the length of a register-interval is usually limited by the size of the register working-set, a strand is typically terminated due to unrelated control flow constraints, and as a result, the strand's register working-set is often smaller than the available register file cache space. We conclude that using register-intervals to place the prefetch operations is essential for \titleShort{} performance.

\subsection{Impact of Number of Warps per SM}
\begin{figure}[h!]
\centering
\includegraphics[trim=15mm 100mm 15mm 100mm,width=0.7\linewidth]{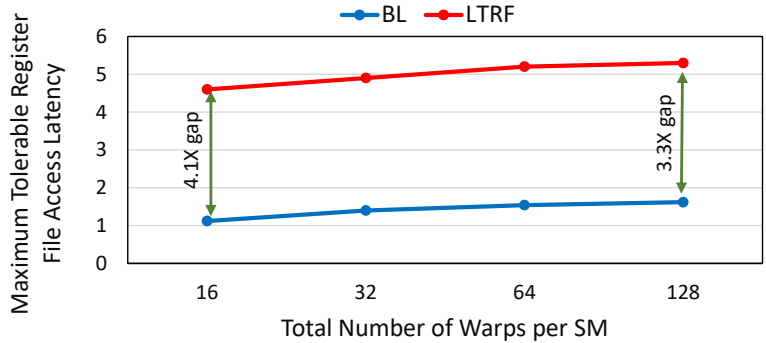}
\caption{Effect of changing number of warps per SM on tolerating register file access latencies.}
\label{fig:change-warp}
\end{figure}

We have already studied the impact of changing the number of active warps on \titleShort{}'s effectiveness in \S~\ref{sec:sensitivity}. In this section, we study how \titleShort{} works when we change the total number of warps per SM. To this end, we compare "the maximum tolerable register file access latency" metric for \titleShort{} and BL designs using various numbers of warps per SM. We consider four different numbers of warps per SM in this experiment: 16, 32, 64, and 128. Figure~\ref{fig:change-warp} reports the results. We make two key observations. First, \titleShort{} improvement compared to BL is higher when we use lower numbers of warps per SM. This is due to the fact that the BL design significantly loses its ability to tolerate register file access latencies when the number of warps per SM is low. However, \titleShort{} still can work with low number of warps, since it relies on register file caching and register prefetching in addition to thread-level parallelism to tolerate register file access latency. Second, we observe almost the same results when we increase the number of warps from 64 to 128, indicating potential saturation of latency tolerance as the number of warps per SM increases. We conclude that \titleShort{} is effective using various numbers of warps.

\section{Related Work}
\label{sec:RW}
To our knowledge, this paper is the first to design a latency-tolerant register file architecture for GPUs by (1) prefetching the entire register working set of a warp from the main register file to the register file cache using the notion of register-intervals, and (2) overlapping the prefetch latency with the execution of other active warps. \titleShort{} opens a window for many optimizations in the main register file that greatly increase the effective capacity at the expense of higher access latency. We have already compared \titleShort{} extensively to various hardware and software register caching proposals for GPUs~\cite{cache1,cache2} in \S~\ref{sec:eval}. In this section, we describe other related work in register file caching and register file scalability. 

\noindent\textbf{Register File Caching.} Few works have explored hardware- and software-managed hierarchical register files for GPUs~\cite{cache1,cache2,asghari-micro-2020}. These works focus on other objectives, such as energy efficiency, rather than latency-tolerance, and expose the higher latencies of slow register files to the execution. Regless~\cite{regless} is a concurrent work that slices the computation graph into regions and allocates operand storage for the regions to replace the register file with a small operand staging unit. However, Regless targets power reduction rather than latency tolerance as the main objective.

Register file caching and hierarchical register files have been widely investigated for CPU architectures. Most of those works focus on superscalar or VLIW processors~\cite{RFC_CPU3,RFC_CPU4,RFC_CPU5,RFC_CPU6,RFC_CPU7,RFC_CPU8,RFC_CPU10,RFC_CPU13}. Such architectures are often able to hide the larger access latency of the slower register file levels via instruction level parallelism. As a result, the main focus of this line of work has been on efficient ways of integrating hierarchical register files into deep out-of-order pipelines and orchestrating the interactions between rename/issue mechanisms and register movements among different levels~\cite{RFC_CPU6, RFC_CPU7}. However, these techniques are usually not applicable to GPUs as GPUs have limited support for instruction-level parallelism. Another line of work focuses on software-managed hierarchical files with different ISA-visible register banks that have different sizes and speeds~\cite{RFC_CPU2,RFC_CPU11,RFC_CPU12,RFC_CPU9} where the compiler orchestrates register placement and movement. The CRAY-1 system~\cite{RFC_CPU1} is an example architecture that implements a compiler-controlled hierarchical register file where software instructions explicitly manage the register movement between the two levels. Such techniques are suitable mainly for VLIW/vector processors and are not effective when used with GPUs where dynamic thread interleavings are unknown at compile-time as the GPU compiler is not able to schedule register movements to overlap them with the execution of other threads.

\noindent\textbf{Register File Scalability.} There are many techniques that improve the scalability or efficiency of the register files. These techniques employ dimming and power-gating~\cite{gate1,gate3}, compression~\cite{RF_comp}, new memory technologies~\cite{sttram1, sttram2, sttram3, edram1, edram2, edram3, edram4, DWM_RF, TFET, murali_RF}, and virtualization~\cite{virtual_RF1,Zorua,CORF,khorasani2018regmutex}. All these techniques likely cause an increase in register file access latency. \titleShort{} can be synergistically combined with these techniques and can enable them to tolerate the long register file access latencies. Hence, we believe \titleShort{} is a substrate that enables optimizations in GPU register files, which might otherwise not always be desirable, efficient, or high performance.

\noindent\textbf{Prefetching.}
Data prefetching techniques in memory systems aim to improve the overall performance by fetching data from the lower levels of the memory hierarchy to the higher levels ahead of its first access~\cite{fetch3,lee2011prefetch,bera2019dspatch,vijaykumar2018case,srinath2007feedback,vijaykumar2018locality,ebrahimi2009coordinated,lee2009improving,wenisch2010making, 10.1145/2000064.2000081, li2019framework, lai2001dead, somogyi2006spatial, somogyi2009spatio, ferdman2011proactive, grimsrud1993multiple, orosa2018avpp, liu2014flap, cassell2018disk, jiang2013prefetching, ding2007diskseen, baek2008prefetching, cao1996implementation, patterson1995informed, vandebogart2009reducing, summers2014automated, son2006energy, song2007energy, brown2002speculative, chappell1999simultaneous, zhang2007accelerating, collins2001speculative, kim2004physical, kim2002design, liao2002post, lu2005dynamic, luk2001tolerating, sundaramoorthy2000slipstream, zilles2001execution, solihin2002using, annavaram2001data, collins2001dynamic, kamruzzaman2011inter, atta2015self, ibrahim2003slipstream, chilimbi2002dynamic, Bakhsh-1, Bakhsh-2, Bakhsh-3, Bakhsh-4, Bakhsh-5, SW_prefetcher3, roth1999effective, lipasti1995spaid, fuchs2014loop, lu2003performance, SW_prefetch2, chidambaram2012application, SW_prefetch, gummaraju2005stream, kim2016path, kadjo2014b, michaud2016best, pugsley2014sandbox, yu2015imp, ishii2009access, chen1995effective, jouppi1990improving, sharif2011data, mutlu2005address, jain2013linearizing,spare-register-aware}. 
Some methods aim to reduce I/O request response time by prefetching data from disk to memory or NAND flash based solid state disks (SSDs) as a second-level disk cache. By analyzing the access pattern of disk requests, these methods predict the stream of blocks that will be accessed in the future~\cite{grimsrud1993multiple, liu2014flap, cassell2018disk, jiang2013prefetching, ding2007diskseen, baek2008prefetching, cao1996implementation, patterson1995informed, vandebogart2009reducing, summers2014automated, son2006energy, song2007energy}.

Some prefetching methods prefetch cache lines to different levels of caches~\cite{mutlu_runahead,mutlu2003runahead,vijaykumar2018case,ahn2015scalable,ibrahim2003slipstream, kim2004physical, kim2002design, liao2002post, solihin2002using, annavaram2001data, collins2001dynamic, kamruzzaman2011inter, Bakhsh-1, Bakhsh-2, Bakhsh-3, Bakhsh-4, Bakhsh-5, atta2015self, brown2002speculative, ebrahimi2009techniques, chappell1999simultaneous, zhang2007accelerating, collins2001speculative, lu2005dynamic, luk2001tolerating, sundaramoorthy2000slipstream, zilles2001execution, chilimbi2002dynamic, SW_prefetch, SW_prefetcher3, roth1999effective, lipasti1995spaid, fuchs2014loop, lu2003performance, SW_prefetch2, chidambaram2012application, gummaraju2005stream, kim2016path, kadjo2014b, michaud2016best, pugsley2014sandbox, yu2015imp, ishii2009access, chen1995effective, jouppi1990improving, sharif2011data, mutlu2005address, jain2013linearizing, mutlu2005techniques,mutlu2006efficient,hashemi2016continuous,hashemi2015filtered,ramirez2010efficient,mutlu2005reusing,mutlu2005analysis}, including thread-based prefetching~\cite{mutlu2005analysis,mutlu2005reusing,ramirez2010efficient,hashemi2015filtered,hashemi2016continuous,mutlu_runahead,mutlu2003runahead,brown2002speculative, chappell1999simultaneous, zhang2007accelerating, collins2001speculative, ibrahim2003slipstream, kim2004physical, kim2002design, liao2002post, lu2005dynamic, luk2001tolerating, sundaramoorthy2000slipstream, zilles2001execution, solihin2002using, annavaram2001data, collins2001dynamic, kamruzzaman2011inter, atta2015self,mutlu2005techniques,mutlu2006efficient}, software-based prefetching~\cite{brown2002speculative, ebrahimi2009techniques, ibrahim2003slipstream, chilimbi2002dynamic, SW_prefetch, SW_prefetcher3, roth1999effective, lipasti1995spaid, fuchs2014loop, lu2003performance, SW_prefetch2, chidambaram2012application, gummaraju2005stream}, and hardware-based prefetching~\cite{kim2016path, kadjo2014b, michaud2016best, pugsley2014sandbox, yu2015imp, ishii2009access, chen1995effective, jouppi1990improving, sharif2011data, mutlu2005address, jain2013linearizing,spare-register-aware,bera2019dspatch}. Thread-based prefetching techniques~\cite{brown2002speculative, chappell1999simultaneous, zhang2007accelerating, collins2001speculative, ibrahim2003slipstream, kim2004physical, kim2002design, liao2002post, lu2005dynamic, luk2001tolerating, sundaramoorthy2000slipstream, zilles2001execution, solihin2002using, annavaram2001data, collins2001dynamic, kamruzzaman2011inter, atta2015self} execute special threads, known as prefetcher or run-ahead threads, to prefetch data. The downside of thread-based prefetching is that the processor should have enough extra resources to execute the prefetching threads. Software-based prefetching techniques~\cite{ibrahim2003slipstream, chilimbi2002dynamic, SW_prefetcher3, roth1999effective, lipasti1995spaid, fuchs2014loop, lu2003performance, SW_prefetch2, chidambaram2012application, SW_prefetch, gummaraju2005stream} receive hints from the compiler or the programmer to prefetch data. However, it is difficult to detect all access patterns in the programs that have complex and unpredictable access patterns, such as server and scale-out applications. Hardware-based prefetching techniques ~\cite{kim2016path, kadjo2014b, michaud2016best, pugsley2014sandbox, yu2015imp, ishii2009access, chen1995effective, jouppi1990improving, sharif2011data, mutlu2005address, jain2013linearizing} exploit a dedicated hardware mechanism in the processor to predict access patterns dynamically. Hardware-based methods are able to predict more complex addresses in comparison with software-based prefetching techniques, especially for servers and scale-out applications that perform a large amount of pointer-chasing, at the price of higher hardware cost.

\section{Conclusion}
\label{sec:conclusion}
We propose \titleShort{}, a new \emph{latency-tolerant} hierarchical register file design for GPUs. The key mechanism of \titleShort{} is a near-perfect register prefetching scheme that divides the application control flow graph into register-intervals and brings the entire register working set of a warp from the main register file to the register cache at the beginning of each register-interval. As a result, a warp experiences the fast register cache access latency, rather than the long access latency of the large main register file. We devise a compile-time register renumbering technique on top of \titleShort{} to resolve register bank conflicts. 
An example evaluation result shows that \titleShort{} combined with register renumbering technique enables us to implement the main register file with emerging high-density high-latency memory technologies, enabling 8$\times$ larger register file capacity and improving overall GPU performance by 34\%. We believe that \titleShort{} paves the way for many power/area optimization techniques in the main register file that likely increase the register access latency. We conclude that, by making the performance impact of a slower register file more tolerable, \titleShort{} enables a large design space to architects, where tradeoffs between power, area, and latency of the register file can be explored more freely to optimize system-level goals.

\bibliographystyle{ACM-Reference-Format}
\bibliography{main}

\end{document}